\documentclass{IEEEtran}
\usepackage{graphicx}  
\usepackage{dcolumn}   
\usepackage{bm}        
\usepackage{verbatim}   
\usepackage{graphicx}
\usepackage{epstopdf}
\usepackage{array}
\usepackage{cite}
\usepackage{psfrag}
\usepackage{booktabs}
\usepackage{footnote}
\usepackage[most]{tcolorbox}
\usepackage{epsfig}
\usepackage{subfigure}
\usepackage{psfrag}
\usepackage{amssymb}
\usepackage{amsmath,bm}
\usepackage{lipsum}
\usepackage{mathtools}
\usepackage{cuted}
\usepackage[utf8]{inputenc}
\usepackage{lipsum} 

\newcolumntype{P}[1]{>{\centering\arraybackslash}p{#1}}

\newcommand{\ve}[1]{\boldsymbol{#1}}
\newcommand{\te}[1]{\overline{\overline{#1}}}
\usepackage[framemethod=TikZ]{mdframed}
\usepackage{lipsum}
	
\begin{document}	
\title{Four-Dimensional Wave Transformations By Space-Time Metasurfaces}
\author{Sajjad Taravati and George V. Eleftheriades, \IEEEmembership{Fellow, IEEE}
	\thanks{}
		\thanks{Sajjad Taravati and George V. Eleftheriades are with the Edward S. Rogers Sr. Department of Electrical and Computer Engineering, University of Toronto, Toronto, Ontario M5S 3H7, Canada. 
		
		Email: sajjad.taravati@utoronto.ca.
		}}

\markboth{}%
{*** \MakeLowercase{\textit{et al.}}: Bare Demo of IEEEtran.cls for IEEE Journals}

\maketitle	
\begin{abstract}
Static metasurfaces have shown to be prominent compact structures for reciprocal and frequency-invariant transformation of electromagnetic waves in space. However, incorporating temporal variation to static metasurfaces would result in dynamic apparatuses which are capable of four-dimensional tailoring of both the spatial and temporal characteristics of electromagnetic waves, leading to functionalities that are far beyond the capabilities of conventional static metasurfaces. This includes nonreciprocal full-duplex wave transmission, pure frequency conversion, parametric wave amplification, spatiotemporal decomposition, and space-time wave diffraction. This paper reviews recent progress and opportunities offered by space-time metasurfaces to break reciprocity, revealing their potential for low-energy, compact, integrated non-reciprocal devices and sub-systems, and discusses the future of this exciting research front.
\end{abstract}

	
\section{Introduction}
Controlled transformation of electromagnetic fields has advanced drastically in recent years thanks to the advent and evolution of metamaterials and metasurfaces~\cite{ebbesen1998extraordinary,pendry2006controlling,Capasso_Sc_2011,kang2012wave,ni2012broadband,Taravati_Kishk_MicMag_2019}. Three-dimensional static metamaterials and metasurfaces has made a substantial progress in wave engineering applications~\cite{elachi1976waves,shelby2001experimental,pendry2006controlling,Capasso_Sc_2011,cheng2014wave,celli2014low}. However, recently there has been a growing interest on four-dimensional metasurfaces, where adding the temporal variation to three-dimensional metasurfaces leads to functionalities that are far beyond the capabilities of conventional static metasurfaces. For instance, asymmetric wave transmission
may be achieved by spatially asymmetric structures when
multiple modes are involved at different ports, whereas nonreciprocal wave transmission is a noticeably more challenging task
that requires an external field for breaking the time-invariance of the structure, biasing with odd-symmetric quantities under time reversal, or nonlinear materials. Among these nonreciprocity approaches, space-time (ST) modulation is of high interest thanks to its immense capability for affecting the spectrum of the electromagnetic waves while breaking the time reversal symmetry. 

ST metasurfaces provide huge degrees of freedom for arbitrary alteration of the wavevector and temporal frequency of electromagnetic waves, leading to an advanced four-dimensional wave processing from acoustics and microwaves~\cite{Alu_PRB_2015,Taravati_2017_NR_Nongyro,zhang2018space,Taravati_Kishk_PRB_2018,zhang2018space,Grbic2019serrodyne,Taravati_Kishk_TAP_2019,zang2019nonreciprocal_metas,taravati_PRApp_2019,saikia2019frequency,ramaccia2019phase,taravati2020full,Taravati_MixMet_APS_2020} to terahertz and
optics~\cite{Fan_APL_2016,Salary_2018,guo2019nonreciprocal,salary2019dynamically,salary2020time,sedeh2020time}. They represent a class of compact \textit{dynamic wave processors}, which have been recently proposed for extraordinary transmission of electromagnetic waves. Such four-dimensional compact apparatuses are endowed with unique properties not readily seen in conventional static metamaterials and metasurfaces. ST metasurfaces may take advantage of space-time modulation capabilities, including nonreciprocal frequency generation~\cite{Taravati_PRB_Mixer_2018,Grbic2019serrodyne,saikia2019frequency,chu2020controllable}, parametric wave amplification~\cite{Cullen_NAT_1958,Tien_JAP_1958,Oliner_PIEEE_1963,Chu1972,elachi1972electromagnetic,Taravati_2017_NR_Nongyro,pendry2020new,zhu2020tunable}, asymmetric dispersion~\cite{Taravati_PRB_2017,Milford_Elnaggar_2017,Taravati_PRAp_2018}, and energy accumulation~\cite{lurie2017energy}. Frequency generation and directivity are of particular interest in space-time-modulated (STM) slabs~\cite{Taravati_PRB_2017,Taravati_thesis,Taravati_PRB_Mixer_2018,Taravati_PRAp_2018,saikia2019frequency,barati2020topological,sedeh2020time,huidobro2019fresnel,Taravati_MixMet_APS_2020}, which are endowed by asymmetric periodic electromagnetic transitions in their dispersion diagram~\cite{Taravati_PRB_2017,Milford_Elnaggar_2017,Taravati_thesis,Taravati_Kishk_TAP_2019}. In practice, the ST modulation is achieved through pumping the external energy into the medium~\cite{Taravati_PRB_2017,Taravati_PRB_Mixer_2018,Taravati_Kishk_PRB_2018,Grbic2019serrodyne,taravati2020full}. 

Some of the recently proposed applications of STM metamaterials and metasurfaces include mixer-duplexer-antenna~\cite{Taravati_LWA_2017}, unidirectional beam splitters~\cite{Taravati_Kishk_PRB_2018}, nonreciprocal filters~\cite{alvarez2019coupling,wu2019isolating}, signal coding metasurfaces~\cite{taravati_PRApp_2019,zhang2019dynamically}, ST metasurfaces for advanced wave engineering and extraordinary control over electromagnetic waves~\cite{Alu_PRB_2015,Fan_APL_2016,stewart2017finite,Salary_2018,Taravati_Kishk_TAP_2019,zang2019nonreciprocal_metas,inampudi2019rigorous,elnaggar2019generalized,wang2018photonic,Grbic2019serrodyne,Taravati_Kishk_MicMag_2019,ptitcyn2019time,du2019simulation,wang2019multifunctional,taravati2020full,salary2019dynamically,li2020time}, nonreciprocal platforms~\cite{Taravati_thesis,wentz1966nonreciprocal,Taravati_PRB_2017,Taravati_PRB_SB_2017,Taravati_PRAp_2018,oudich2019space,chegnizadeh2020non,li2020time}, frequency converters~\cite{Taravati_PRB_Mixer_2018,Grbic2019serrodyne,saikia2019frequency,chu2020controllable,Taravati_MixMet_APS_2020}, time-modulated antennas~\cite{shanks1961new,ramaccia2018nonreciprocity,taravati2018space,zang2019nonreciprocal} spectral camouflage metasurfaces~\cite{liu2019time}, antenna-mixer-amplifiers~\cite{Taravati_AMA_PRApp_2020}, and 
enhanced resolution imaging photonic crystals~\cite{manzoor2020enhanced}. This strong capability of STM media is due to their unique interactions with the incident field~\cite{Taravati_PRB_2017,li2019nonreciprocal,liu2018huygens,du2019simulation,elnaggar2019modelling,taravati2019_mix_ant}. 

This paper provides a review on the properties of ST metasurfaces, their analysis, and their potential applications in modern and future wireless communication systems, and wave tailoring processors.
We first present key properties of ST interfaces, including spatial interfaces, temporal interfaces, and ST interfaces. Then, we show that a nonreciprocal metasurface acts as a very thin ST slab, i.e., a moving metasurface. Next, analysis of general STM metasurfaces will be given, including derivation of scattered electromagnetic fields, four-dimensional dispersion diagrams, boundary conditions, and spatiotemporal decomposition. Then, full-wave finite-difference time domain (FDTD) simulation of ST metasurfaces are presented.

To showcase various applications of ST metasurfaces, we consider both transmissive and reflective ST  diffraction gratings, and discuss their nonreciprocal and asymmetric response. We also discuss a ST diffraction-code multiple access system, a unidirectional beam splitter based on superluminal ST modulation, a nonreciprocal-beam-steering metasurface, and an antenna-mixer-amplifier transceiver metasurface.

\section{Electromagnetic Waves in Space-Time}
\begin{figure*}
	\begin{center}
	\centering{\includegraphics[width=2\columnwidth]{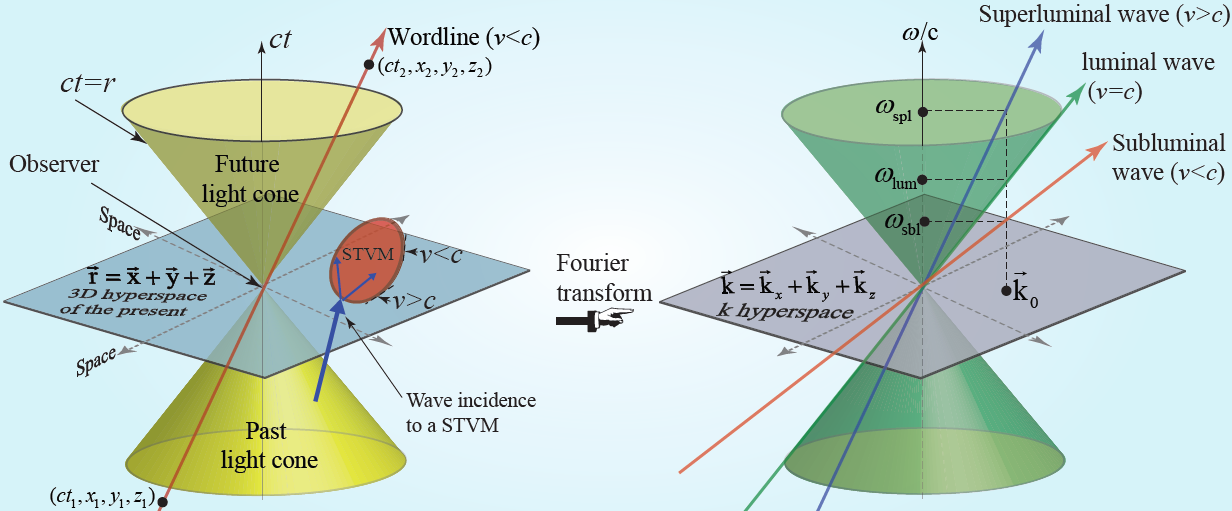}}  
	\caption{Space-time (ST) Fourier pair diagrams.}
	\label{Fig:ST}
\end{center}
\end{figure*}

Figure~\ref{Fig:ST} shows the Minkowski ST diagram and its Fourier transformed pair known as the dispersion diagram. The four-dimensional Minkowski ST diagram includes two light cones, representing propagation of the light in the past and future. The two cones have their apexes at the present, where the three-dimensional ($x$, $y$ and $z$) hyperspace exists. Any discontinuity in the four dimensional ST diagram may result in forward and backward waves in space. Analysis and design of ST media and metasurfaces can be substantially eased by understanding the Fourier pair of the Minkowski diagram. To best investigate the wave diffraction by a STM grating, we first study the interaction of the electromagnetic wave with space and time interfaces, separately. In general, three different ST discontinuities may be studied as follows.

\subsection{Space Interface}
Figure~\ref{Fig:boundary}(a) sketches the ST diagram of a spatial interface between two media of refractive indices $n_1$ and $n_2$, respectively, in the plane ($z,ct$). This figure shows scattering of forward and backward fields and conservation of energy and momentum for different scenarios. The temporal axis of the Minkowski ST diagram is scaled with the speed of light $c$, and therefore is labeled by $ct$ for changing the dimension of the addressed physical quantity from time to length, in accordance to the dimension associated to the spatial axes labeled $z$. This problem represents the textbook case of electromagnetic wave incidence and scattering from a spatial (static) interface, where $n(z<0)=n_1$ and $n(z>0)=n_2$. The boundary conditions are derived by applying the fundamental physical fact that all physical quantities must remain bounded
everywhere and at every time to the space and time derivatives in sourceless Maxwell equations
\begin{equation}
\nabla \times \textbf{E}=- \dfrac{\partial \textbf{B}}{\partial t} \quad \text{and} \quad \nabla \times \textbf{H}= \dfrac{\partial \textbf{D}}{\partial t}.
\end{equation}

The discontinuity of the tangential components of electric and magnetic fields at $z=z_0$ would result in unbounded and singular electromagnetic fields at the interface, which is not physical. Therefore, the tangential components of the electric and magnetic fields must be continuous at a space discontinuity, i.e.,
\begin{equation}
\hat{z} \times (\textbf{E}_2-\textbf{E}_1)|_{z=z_0}=0 \quad \text{and} \quad \hat{z} \times (\textbf{H}_2-\textbf{H}_1)|_{z=z_0}=0. 
\end{equation}

As a result, the wavenumber $k$ changes, i.e., energy is preserved but momentum changes, such that the forward transmitted wave in the region 2 corresponds to $k_\text{t}= k_\text{i} n_2/n_1$, whereas the temporal frequency of the transmitted wave in region 2 is equal to that of region 1, i.e., $\omega_\text{t}=\omega_\text{i}$.
\begin{subequations}
	\label{Eq:sp_bound}
	\begin{equation}
E_1=\left(e^{ik_\text{i} z}+ R e^{-ik_\text{i} z} \right) e^{-i \omega_\text{i} t},
\quad \text{and} \quad
E_2=T e^{ik_\text{t} z}  e^{-i \omega_\text{i} t},
	\end{equation}
where $R$ and $T$ represent the spatial reflection and transmission coefficients, and defined as
\begin{equation}
R=\dfrac{n_1-n_2}{n_1+n_2} , \quad \text{and} \quad T=\dfrac{2 n_2}{n_1+n_2}.
\end{equation}
\end{subequations}

\subsection{Time Interface}
Figure~\ref{Fig:boundary}(b) shows the ST diagram of a time interface between two media of refractive indices $n_1$ and $n_2$, which is the dual case of the spatial metasurface in Fig.~\ref{Fig:boundary}(a)~\cite{felsen1970wave,biancalana2007dynamics,bacot2016time}. Here, the refractive index suddenly changes from one value ($n_1$) to another ($n_2$) at a given time throughout all space, i.e., $n(t<0)=n_1$ and $n(t>0)=n_2$. The temporal change of the refractive index produces both reflected (backward) and transmitted (forward) waves, which is analogous to the reflected and transmitted waves
produced at the spatial interface between two different media in Fig.~\ref{Fig:boundary}(a). The discontinuity of $\textbf{D}$ and $\textbf{B}$ at $ct=ct_0$ would result in unbounded and singular $\textbf{E}$ and $\textbf{H}$ at the interface, which is not physical. Therefore, $\textbf{D}$ and $\textbf{B}$ must be continuous at a time interface, that is,
\begin{equation}
 (\textbf{D}_2-\textbf{D}_1)|_{ct=ct_0}=0 \quad \text{and} \quad  (\textbf{B}_2-\textbf{B}_1)|_{ct=ct_0}=0.
\end{equation}

The total charge $Q$ and the total flux $\psi$ must remain constant at the moment of the jump from $n_1$ to $n_2$, implying that both transversal and normal components of $\textbf{D}$ and $\textbf{B}$ do not change instantaneously~\cite{morgenthaler1958velocity,fante1971transmission}, which is different than the static case (shown in Fig.~\ref{Fig:boundary}(a)) where only the normal components of the magnetic field $\textbf{B}$ and electric field displacement $\textbf{D}$ are conserved. Specifically, at a time interface, the magnetic field $\textbf{B}$, the electric field displacement $\textbf{D}$ and the wavenumber $k$ are preserved. This yields a change in the temporal frequency of the incident wave so that the frequency of the forward transmitted wave in region 2 corresponds to $\omega_\text{t}= \omega_\text{i} n_1/n_2$, i.e., where momentum is preserved but energy changes.
\begin{subequations}
	\label{Eq:time_bound}
	\begin{equation}
D_1=  e^{ik_\text{i} z}  e^{-i \omega_\text{i} t}
\quad \text{and} \quad
D_2= e^{i k_\text{i} z} \left(\widehat{T} e^{-i \omega_\text{t} t} +\widehat{R} e^{i \omega_\text{t} t}  \right),
	\end{equation}
where $\widehat{R}$ and $\widehat{T}$ represent the temporal reflection and transmission coefficients, and defined as
\begin{equation}
\widehat{R}=\dfrac{n_1}{n_2} \dfrac{\eta_1-\eta_2}{2\eta_1}, \quad \text{and} \quad \widehat{T}=\dfrac{n_1}{n_2} \dfrac{\eta_1+\eta_2}{2\eta_1}.
\end{equation}
\end{subequations}

where $\eta_1=\sqrt{\mu_1/\epsilon_1}$ and $\eta_2=\sqrt{\mu_2/\epsilon_2}$.

\subsection{Space-Time Interface}
\begin{figure}
	\begin{center}
		\includegraphics[width=1\columnwidth]{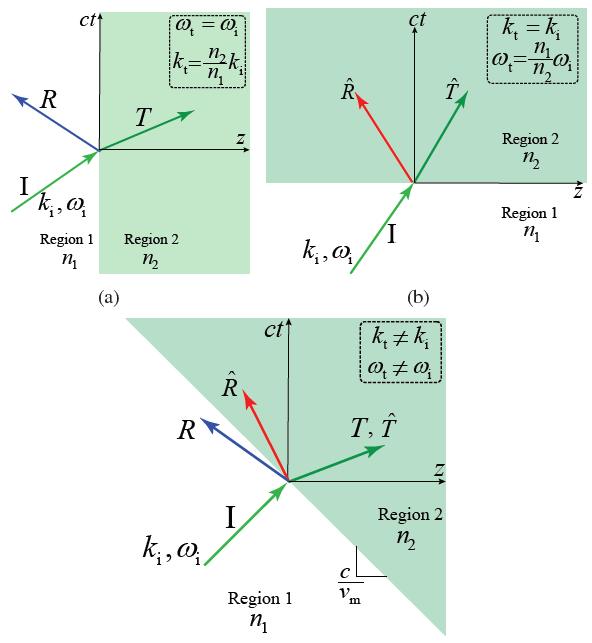}
		\caption{ST diagrams showing scattering of forward and backward fields 
			and conservation of energy and momentum for different scenarios. (a) Spatial interface, i.e., $n(z<0)=n_1$ and $n(z>0)=n_2$. (b) Temporal interface, i.e., $n(t<0)=n_1$ and $n(t>0)=n_2$. (c) ST interface, i.e., $n(z/c+t<0)=n_1$ and $n(z/c+t>0)=n_2$.}
		\label{Fig:boundary}
	\end{center}
\end{figure}

Figure~\ref{Fig:boundary}(c) depicts the ST diagram of a ST interface, i.e., $n(z/c+t<0)=n_1$ and $n(z/c+t>0)=n_2$, as the combination of the space and time interfaces in Figs.~\ref{Fig:boundary}(a) and~\ref{Fig:boundary}(b), respectively. It may be seen that the ST interface resembles the spatial interface configuration in Fig.~\ref{Fig:boundary}(a) in the region $n=n_1$ and the temporal interface configuration in Fig.~\ref{Fig:boundary}(b) for $n=n_2$. 

The reflection and transmission from a subluminal ST interface reads
\begin{subequations}
	\label{Eq:ST_sub}
\begin{equation}
R=\dfrac{\eta_2-\eta_1}{\eta_1+\eta_2} \dfrac{v_1-v_\text{m} }{v_1+v_\text{m} }
\quad \text{and} \quad
T=\dfrac{2\eta_2}{\eta_1+\eta_2} \dfrac{v_1-v_\text{m}}{v_2-v_\text{m} },
\end{equation}
and the temporal and spatial frequencies of the reflected and transmitted waves read	
\begin{equation}	
\omega_R=\omega_\text{i} \dfrac{v_1-v_\text{m} }{v_1+v_\text{m} },
\quad \text{and} \quad	
\omega_T=\omega_\text{i} \dfrac{v_1-v_\text{m}}{v_2-v_\text{m} },
\end{equation}	
\begin{equation}	
k_R=k_\text{i} \dfrac{v_1-v_\text{m} }{v_1+v_\text{m} },
\quad \text{and} \quad	
k_T= k_\text{i} \dfrac{v_1-v_\text{m}}{v_2-v_\text{m} },
\end{equation}	
\end{subequations}
where $v_1=c/n_1$ and $v_2=c/n_2$. 

The reflection and transmission from a superluminal ST interface read 
\begin{subequations}
	\label{Eq:ST_sup}
\begin{equation}
\widehat{R}=\dfrac{\eta_2-\eta_1}{2 \eta_1} \dfrac{v_\text{m}-v_1}{v_\text{m}+v_2 }, \quad \text{and} \quad 
\widehat{T}=\dfrac{\eta_1+\eta_2}{2 \eta_1} \dfrac{v_\text{m}-v_1 }{v_\text{m}-v_2 },
\end{equation}
and the temporal and spatial frequencies of the reflected and transmitted waves read
\begin{equation}
\omega_{\widehat{R}}=\omega_\text{i} \dfrac{ v_\text{m}-v_1}{v_\text{m}+ v_2},
\quad \text{and} \quad	
\omega_{\widehat{T}}=\omega_\text{i} \dfrac{v_\text{m}-v_1 }{v_\text{m}-v_2},
\end{equation}
\begin{equation}	
k_{\widehat{R}}=k_\text{i} \dfrac{v_\text{m}-v_1 }{v_\text{m}+v_2 },
\quad \text{and} \quad	
k_{\widehat{T}}= k_\text{i} 
\dfrac{v_\text{m}-v_1}{v_\text{m}-v_2}.
\end{equation}
\end{subequations}

The pure-space interface is the $v_\text{m}=0$ limit of a subluminal interface, while the pure-time interface is the $v_\text{m}=\infty$ limit of a superluminal interface.

\begin{figure}
		\begin{center}
	\includegraphics[width=1\columnwidth]{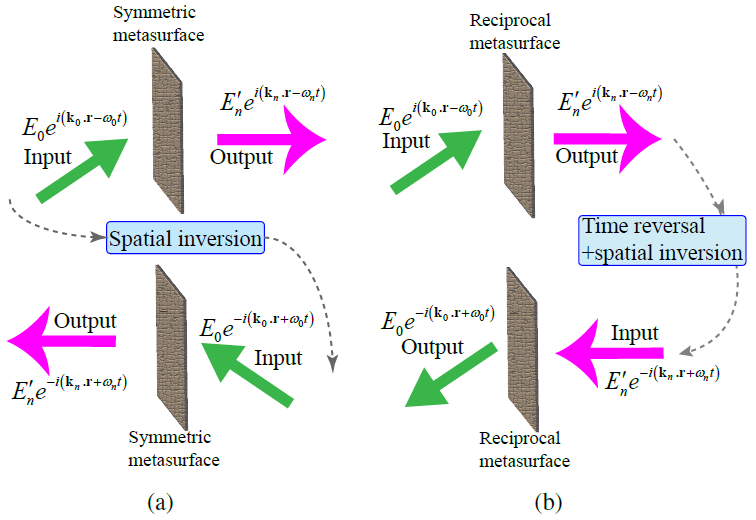} 
	\caption{Schematic of the experimental set-up configurations for validation of symmetric and reciprocal response of electromagnetic systems. (a) The electromagnetic symmetry of the system is validated, in which the backward problem is the \textit{spatial inversion} of the forward problem. (b) The electromagnetic reciprocity of the system is validated, in which the backward problem is the \textit{spatial inversion of the time-reversed} forward problem.} 
	\label{fig:3}
	\end{center}
\end{figure}

The difference between the excitation and response for validation of the symmetry and reciprocity of electromagnetic systems, associated with new frequency generation, is clarified in Figs.~\ref{fig:3}(a) and~\ref{fig:3}(b). Figure~\ref{fig:3}(a) shows the forward and backward problems for the symmetry test of a particular symmetric electromagnetic system, where the backward problem is represented by the spatial inversion of the forward problem, i.e., the applied excitation wave (input) of the backward problem must be the spatial inversion of the excitation wave (input) of the forward problem. As a result, for a symmetric system, the output of the backward problem would be exactly the spatial inversion of the output of the forward problem. Otherwise, the system is asymmetric. Figure~\ref{fig:3}(b) shows the forward and backward problems for the reciprocity test of a particular reciprocal electromagnetic system, where the backward problem is the spatial inversion of the \textit{time-reversed} of the forward problem, i.e., the applied excitation wave (input) of the backward problem must be the spatial inversion of the output of the forward problem. As a result, for a reciprocal system, the output of the backward problem would be exactly the spatial inversion of the input of the forward problem. Otherwise, the system is nonreciprocal.

\section{Experimental Demonstration of ST Interface}
\begin{figure}
\begin{center}
		\includegraphics[width=0.8\columnwidth]{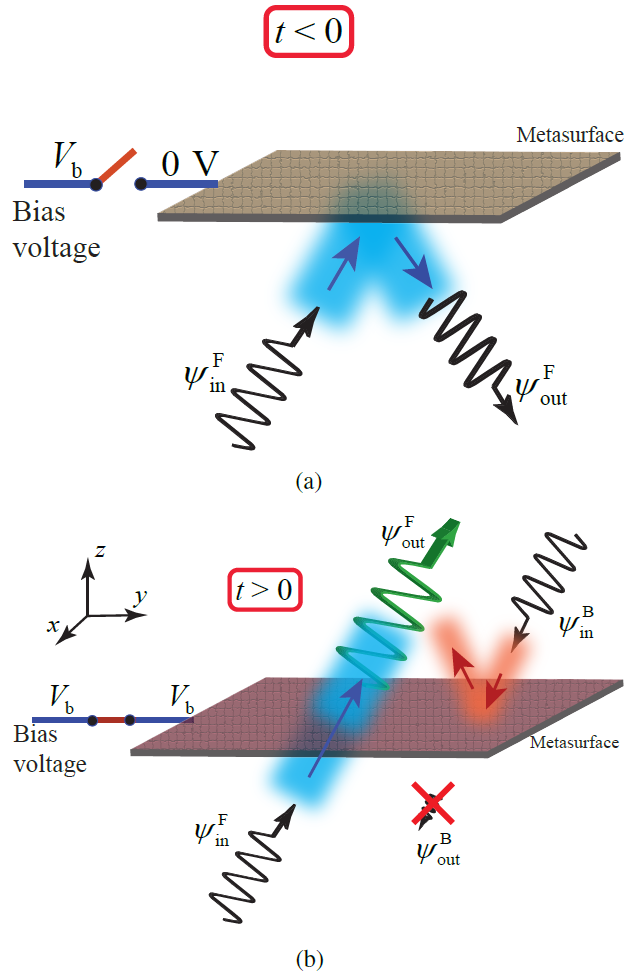}
		\caption{Nonreciprocal nongyrotropic metasurface. (a) For $t>0$ operating as a reflective sheet. (b) For $t>0$ operating as a nonreciprocal transmissive sheet.}
		\label{Fig:4}
		\end{center} 
\end{figure}

Figures~\ref{Fig:4}(a) and~\ref{Fig:4}(b) depict the operation principle of the nonreciprocal nongyrotropic metasurface. For $t<0$, the metasurface operates as a reflector where a $+z$-direction traveling wave is reflected by the metasurface and travels back along $-z$ direction. For $t>0$, the metasurface operates as a nonreciprocal sheet, where a $+z$-direction traveling wave passes through the metasurface with gain and without polarization alteration, whereas a wave traveling along the opposite direction from is being reflected by the metasurface. The transmission scattering parameters of the metasurface are not equal, i.e., $S_{21} > S_{12}$, where $S_{21}=\ve{\psi} _\text{out}^\text{F}/\ve{\psi}_\text{in}^\text{F} >1$ and $S_{12}=\ve{\psi}_\text{out}^\text{B}/\ve{\psi}_\text{in}^\text{B} <1$. 

\begin{figure}
	\begin{center}
		\includegraphics[width=1\columnwidth]{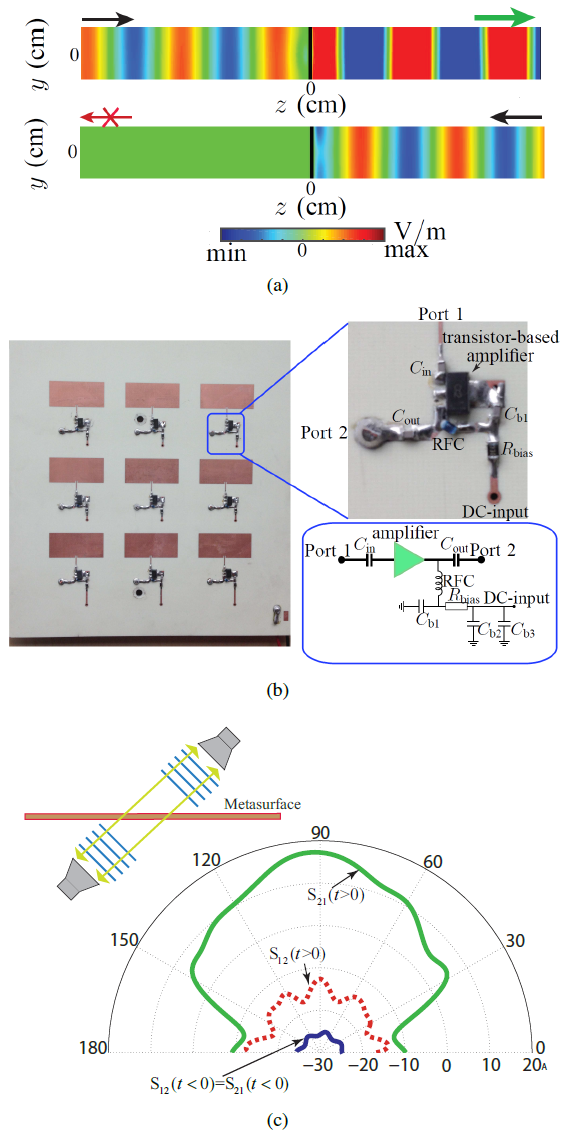}
		\caption{Nonreciprocal metasurface. (a)~Full-wave (FDTD) electric field distribution for excitations from the left and right (bottom)~\cite{Taravati_2017_NR_Nongyro}. (b) An image of the fabricated metasurface~\cite{Taravati_2017_NR_Nongyro}. (c) Experimental scattering parameters versus angle at $f=5.9$~GHz for transmission in a straight line under an oblique angle~\cite{Taravati_2017_NR_Nongyro}.}
		\label{Fig:5}
	\end{center} 
\end{figure}

Figure~\ref{Fig:5}(a) shows the full-wave simulation results, where transmission of waves from left to right is allowed and accompanied by power amplification but the transmission of waves from right to left is prohibited. Figure~\ref{Fig:5}(b) shows an image of the fabricated metasurface~\cite{Taravati_2017_NR_Nongyro}. The metasurface is formed by an array of unit cells. Such unit-cells are constituted of two microstrip patch elements interconnected through a unilateral transistor, introducing transmission gain in one direction and transmission loss in the other direction. Fig.~\ref{Fig:5}(c) shows the measured transmission levels for both directions, and for $t<0$ and $t>0$. The metasurface introduces gain over a bandwidth of about $130^\circ$, i.e., from $\theta=25^\circ$ to $155^\circ$ in the $1\rightarrow 2$ direction, while it introduces attenuation by more than $12$~dB in the $2\rightarrow 1$ direction. This corresponds to an isolation level of more than $21$~dB across the bandwidth. The power gain makes the metasurface particularly efficient as a repeater device. Furthermore, in contrast to other nonreciprocal metasurfaces, the structure is fairly broadband and its bandwidth can be further enhanced by various standard techniques~\cite{Garg_2001}. The operating angular sectors of this metasurface over $100^\circ$, are much greater than those of typical metasurfaces.

\begin{figure}
	\begin{center}
		\includegraphics[width=0.8\columnwidth]{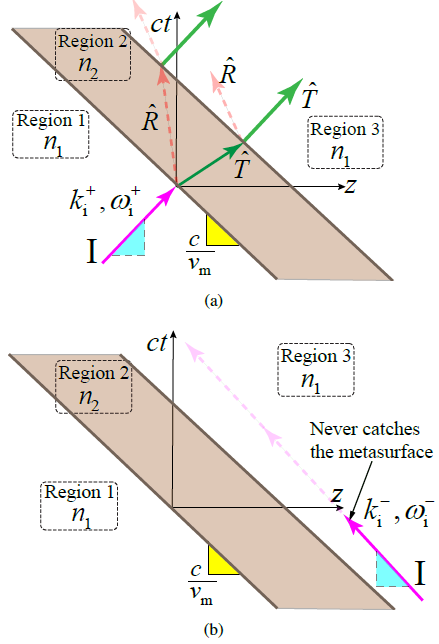}
		\caption{ST representation of the nonreciprocal nongyrotropic metasurface in Figs.~\ref{Fig:4} and~\ref{Fig:5}. (left) Forward wave incidence leading to full transmission, and (right) Backward wave incidence, where the wave can not catch up the metasurface.}
		\label{Fig:6}
	\end{center}
\end{figure}

The nonreciprocal nongyrotropic metasurface in Fig.~\ref{Fig:5} may be represented as a moving metasurface in Fig.~\ref{Fig:6}. The general case of a bianisotropic medium reads~\cite{Taravati_2017_NR_Nongyro}
\begin{subequations}
	\label{Eq:ConstRel}
	\begin{equation}
	\ve{D} = \te{\epsilon} \cdot \ve{E} + \te{\xi}\cdot\ve{H},
	\end{equation}
	\begin{equation}
	\ve{B} =  \te{\zeta}\cdot \ve{E} + \te{\mu}\cdot\ve{H}.
	\end{equation}
\end{subequations}

The continuity equations of a metasurface may be expressed as
\begin{subequations}
	\label{Eq:MSBC}
	\begin{equation}
	\hat{z}\times\Delta\ve{H}
	=j\omega\epsilon_0\te{\chi}_\text{ee}\cdot\ve{E}_\text{av}+jk_0\te{\chi}_\text{em}\cdot\ve{H}_\text{av},
	\end{equation}
	\begin{equation}
	\Delta\ve{E}\times\hat{z}
	=j\omega\mu_0 \te{\chi}_\text{mm}\cdot\ve{H}_\text{av}+jk_0\te{\chi}_\text{me}\cdot\ve{E}_\text{av},
	\end{equation}
\end{subequations}
where $\Delta$ and the subscript `av' represent, respectively, the difference of the fields and the average of the fields between the two sides of the metasurface. Eq.~\eqref{Eq:MSBC} provides a relation between the electromagnetic fields on the two sides of a metasurface and its susceptibilities, in the absence of normal susceptibility components. The constitutive parameters of the metasurface may be represented according to the susceptibilities in~\eqref{Eq:MSBC} as $\te{\epsilon}=\epsilon_0(\te{I}+\te{\chi}_\text{ee})$, $\quad\te{\mu}=\mu_0(\te{I}+\te{\chi}_\text{mm})$, $\te{\xi}=\te{\chi}_\text{em}/c_0$,$\quad\te{\zeta}=\te{\chi}_\text{me}/c_0$.

We then seek for the susceptibilities that provide the nonreciprocal nongyrotropic response of the metasurface by substituting the electromagnetic fields of the corresponding transformation into~\eqref{Eq:MSBC}. Such a transformation includes passing a $+z$-propagating plane wave through the metasurface and complete absorption of a $-z$-propagating plane incident wave, yielding
\begin{subequations}
	\label{eq:ChiTrans}
	\begin{equation}
	\label{eq:ChiTrans1}
	\te{\chi}_\text{ee}= -\frac{j}{k_0}\begin{pmatrix} 1 & 0 \\ 0 & 1 \end{pmatrix},
	\quad
	\te{\chi}_\text{mm}= -\frac{j}{k_0}\begin{pmatrix} 1 & 0 \\ 0 & 1 \end{pmatrix},
	\end{equation}
	\begin{equation}
	\label{eq:ChiTrans2}
	\te{\chi}_\text{em}= \frac{j}{k_0}\begin{pmatrix} 0 & 1 \\ -1 & 0 \end{pmatrix},
	\quad
	\te{\chi}_\text{me}= \frac{j}{k_0}\begin{pmatrix} 0 & -1 \\ 1 & 0 \end{pmatrix},
	\end{equation}
\end{subequations}
This shows that $\te{\chi}_\text{em}$ and $\te{\chi}_\text{me}$ are the ones that contribute to the nonreciprocity of the metasurface. The form of the susceptibility tensors in~\eqref{eq:ChiTrans} is identical to that of a moving uniaxial medium~\cite{Kong_1986}. To an observer in the rest frame of reference, this tensor set transforms to the bianisotropic set, which assumes motion in the $z$-direction and is characterized by 
\begin{equation}
\label{Eq:CTens}
\begin{pmatrix}
\te{\epsilon}  &  \te{\xi} \\
\te{\zeta} &  \te{\mu}
\end{pmatrix}=
\begin{pmatrix}
\epsilon &  0& 0 & 0& \xi & 0 \\
0 &  \epsilon  &0 &  -\xi & 0 & 0 \\
0 &  0  &\epsilon_z &  0 & 0 & 0 \\
0 &  -\xi & 0  & \mu & 0 & 0\\
\xi &  0 & 0  & 0& \mu & 0 \\
0 &  0 & 0  & 0& 0 & \mu_z
\end{pmatrix},
\end{equation}
where the primes indicate the moving frame of reference and $\epsilon_z$ and $\mu_z$ can take arbitrary values. The elements of~\eqref{Eq:CTens} are determined using the Lorentz transform operation~\cite{Kong_1986} $\te{C} = \te{L}_6^{-1} \cdot \te{C}' \cdot \te{L}_6$. The matrices $\te{C}$ and $\te{L}_6$ are respectively expressed by
\begin{equation}
\label{Eq:CMat}
\te{C}=
\begin{pmatrix}
c(\te{\epsilon} - \te{\xi}\cdot \te{\mu}^{-1} \cdot \te{\zeta}) &  \te{\xi}\cdot\te{\mu}^{-1} \\
-\te{\mu}^{-1}\cdot \te{\zeta} &  \te{\mu}^{-1}/c,
\end{pmatrix}
\end{equation}
and~\cite{Taravati_2017_NR_Nongyro}
\begin{equation}
\label{Eq:L6}
\te{L}_6 =\zeta
\begin{pmatrix}
1 &  0 & 0 & 0 &-\beta & 0 \\
0 &  1 & 0 & \beta & 0 & 0 \\
0 &  0 & 1/\zeta & 0  & 0 & 0 \\
0 &  \beta & 0 & 1 & 0 & 0 \\
-\beta &  0 & 0 & 0 & 1 & 0 \\
0 &  0 & 0 & 0 & 0 & 1/\zeta
\end{pmatrix},
\end{equation}

In Eq.~\eqref{Eq:CMat}, $\zeta= 1/\sqrt{1-(v/c)^2}$, with $v$ being the velocity of the medium. We next seek for the moving uniaxial metasurface that is equivalent to the nonreciprocal metasurface. The forward propagating wave  transmits through the moving metasurface while the backward propagating wave would never reach the metasurface, as shown in Fig.~\ref{Fig:6}. Therefore, the backward wave never passes through. For $T\neq 1$, there exist a complex velocity. The fact that this design approach is practically impossible shows that the engineering approach proves the significance of the realized metasurface in Fig.~\ref{Fig:5}(b).

\section{Analysis of Space-Time (ST) Metasurfaces}\label{sec:anal_sol}

\begin{figure}
	\centerline{\includegraphics[width=0.8\columnwidth]{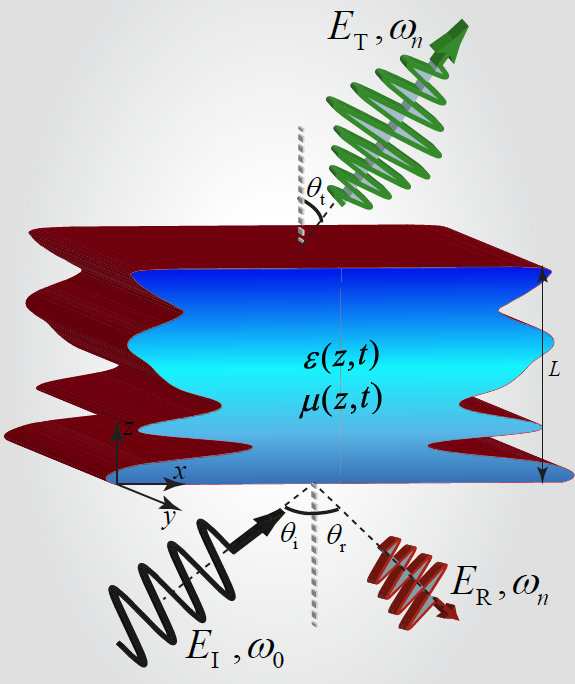}}  
 	\caption{Illustration of oblique incidence and scattering in a general STM metasurface. The metasurface is infinitely extended in the $y$-direction and is hence $y$-invariant~\cite{Taravati_Kishk_TAP_2019}.}
 	\label{Fig:metasurface_concept}
\end{figure}

Consider the STM metasurface in Fig.~\ref{Fig:metasurface_concept}, with the length of $L$, electric permittivity $\epsilon(z,t)$ and magnetic permeability $\mu(z,t)$, sandwiched between two semi-infinite unmodulated media. A general analysis assumes a metasurface with \textit{temporally-periodic} electric permittivity and magnetic permeability, and a general \textit{aperiodic/periodic} spatial variation~\cite{Taravati_Kishk_TAP_2019}. Since the metasurface is time-periodic, its constitutive parameters may be expressed by a time-Floquet series expansion, as
\begin{subequations}\label{eqa:ap_perm}
	\begin{equation}\label{eqa:perm}
	\epsilon(z,t)=\sum_{k=-\infty}^{\infty} \epsilon_\text{$k$,aper}(z) e^{i k\Omega t},
	\end{equation}
	\begin{equation}\label{eqa:ap_permeab}
	\mu(z,t)=\sum_{k=-\infty}^{\infty} \mu_\text{$k$,aper}(z) e^{i k\Omega t}  ,
	\end{equation}
\end{subequations}
where $\Omega$ is the temporal frequency of the modulation, and $\epsilon_\text{$k$,aper}(z)$ and $\mu_\text{$k$,aper}(z)$ are spatially-variant unknown coefficients of the permittivity and permeability, to be determined based on the spatial variation of the metasurface. We consider oblique incidence of a $y$-polarized electric field under an ngle of incidence of $\theta_{\text{i}}$ to the metasurface, as
\begin{equation}\label{eqa:Ei1}
\mathbf{E}_\text{I} (x,z,t)= \mathbf{\hat{y}} E_0 e^{i\left( k_{x,\text{i}} x +k_{z,\text{i}} z- \omega _0 t  \right) },
\end{equation}
where $E_0$ is the amplitude of the incident wave, and $\omega_0$ and $k_\text{I}=\sqrt{k_{x,\text{i}}^2 +k_{z,\text{i}}^2}$ are respectively the temporal and spatial frequencies of the incident wave. Given the temporal periodicity of the metasurface, the electric field inside the metasurface may be represented based on the temporal Bloch-Floquet decomposition as
\begin{subequations}
\begin{equation}\label{eqa:Em1}
\begin{split}
\mathbf{E}_\text{M} (x,z,t)&=\mathbf{\hat{y}} \sum_{n=-\infty}^{\infty} \textbf{E}_{n}(x,z) e^{-i \omega_{n} t},
\end{split}
\end{equation}
where $\omega_{n}=\omega_{0}+n\Omega$, and 
\begin{equation}\label{eqa:Fourier-E}
\textbf{E}_n(x,z)= \textbf{K}_n(z) \sum_{m=-M}^{M} \textbf{E}_{mn} \exp[i(k_{x,\text{i}} x+m\varphi z/L)],
\end{equation}
where $\textbf{E}_{mn}$ is the unknown electric field coefficient matrix, $\varphi$ is the unknown spatial frequency between $0$ and $2\pi$, to be known given the spatial variation of the metasurface, $k_{x,\text{i}}=k_0 \sin(\theta_{\text{i}})=\omega_0 \sqrt{\epsilon_\text{av} \mu_\text{av}}\sin(\theta_{\text{i}})/c$, with $\epsilon_\text{av}$ and $\mu_\text{av}$ being the average permittivity and permeability of the metasurface. In~\eqref{eqa:Fourier-E}, $\textbf{K}_n(z)$ is the principal spatial frequency matrix, given as
\begin{equation}\label{eqa:Fourier-K}
\textbf{K}_n(z)=
\text{diag} \left\lbrace  \left[ \exp(\kappa_{-N} z) \cdots  \exp(\kappa_{0} z) \cdots  \exp(\kappa_{N} z) \right] \right\rbrace 	
\end{equation}
\end{subequations}
where $\kappa_{n}$ are the unknown eigenvalues of the $n$th mode to be found. The magnetic field inside the metasurface reads
\begin{subequations}
\begin{equation}\label{eqa:Hm}
\mathbf{H}_\text{M} (x,z,t)=\dfrac{1}{k_n} \sum_{n=-\infty}^{\infty} \left( -\mathbf{\hat{x}} \beta_{z,n} +\mathbf{\hat{z}} k_{x,\text{i}} \right) \textbf{H}_{n}(x,z) e^{-i \omega_{n} t},
\end{equation}
where $k_n=\sqrt{k_{x,\text{i}}^2+\beta_{z,n}^2}$, and
\begin{equation}\label{eqa:Fourier-H}
\textbf{H}_n(x,z)=\textbf{K}_n(z) \sum_{m=-M}^{M} \textbf{{H}}_{mn} \exp[i(k_{x,\text{i}} x+m\varphi z/L)].
\end{equation}
\end{subequations}

Here, $\beta_{z,n}=\beta_{z,n}(z)=k_{z,\text{i}}+n q(z)$, with $k_{z,\text{i}}=k_0 \cos(\theta_\text{i})$ and $q(z)= \Omega/v_\text{m}(z)=\Omega/\varGamma v_\text{b}(z)$ being the z-component of the incident wavevector and spatial frequency of the modulation, respectively. Here, $v_\text{m}(z)$ and $v_\text{b}(z)$ are phase velocities of the modulation and background medium, respectively, and $\varGamma=v_\text{m}/v_\text{b}$ is the ST velocity ratio~\cite{Taravati_PRB_2017}. The unknown field coefficient matrices, $\textbf{E}_{mn}$, $\textbf{H}_{mn}$, $\beta_{z,n}$ and $\textbf{K}_n$ are to be found through satisfying Maxwell equations, so that
\begin{subequations}
	\begin{equation}\label{eqa:Max1}
	\nabla\times\textbf{E}_\text{M} (x,z,t)=-\dfrac{\partial [\mu(z,t) \textbf{H}_\text{M} (x,z,t)]}{\partial t}
	\end{equation}
	\begin{equation}\label{eqa:Max2}
	\nabla\times\textbf{H}_\text{M} (x,z,t)=\dfrac{\partial [\epsilon (z,t) \textbf{E}_\text{M} (x,z,t)]}{\partial t}
	\end{equation}
\end{subequations}
which may be cast in the form of coupled matrix equations as
\begin{subequations}\label{eqa:coupled_equations}
	\begin{equation}\label{eqa:matrcoup1}
	\textbf{E}(x,z)=\textbf{Z}(z) \textbf{H}(x,z)
	\end{equation}
	\begin{equation}\label{eqa:matrcoup2}
	\textbf{H}(x,z) =\textbf{Y}(z) \textbf{E}(x,z)
	\end{equation}
	where $\textbf{E}(x,z) =[ E_{-N}(x,z) \hdots	E_{0}(x,z) \hdots E_{N}(x,z]^\text{T}$ and $\textbf{H}(x,z) =[H_{-N}(x,z) \hdots H_{0}(x,z) \hdots	H_{N}(x,z)] ^\text{T}$, and 
\begin{equation}\label{eqa:mat-Z}
\textbf{W}(z)=
\begin{bmatrix}
\frac{\omega_{-N}}{k_{-N}} \vartheta_{0}(z) &  \cdots & \frac{\omega_{-N}}{k_{-N}}\vartheta_{2N}(z)\\
\frac{\omega_{-N+1}}{k_{-N+1}}\vartheta_{-1}(z)   &\cdots& \frac{\omega_{-N+1}}{k_{-N+1}}\vartheta_{2N-1}(z)\\
\vdots        &              \ddots           &   \vdots\\
\frac{\omega_{N}}{k_{N}}\vartheta_{-2N}(z)   &\cdots  & \frac{\omega_{N}}{k_{N}} \vartheta_{0}(z)\\
\end{bmatrix}
\end{equation}
\end{subequations}
where $\textbf{W}(z)=\textbf{Z}(z)$ considering $\vartheta_n=\mu_n$ and $\textbf{W}(z)=\textbf{Y}(z)$ considering $\vartheta_n=\epsilon_n$, and where $\omega_{n}=\omega_{0}+n\Omega$. Equations~\eqref{eqa:matrcoup1} and~\eqref{eqa:matrcoup2} form the coupled matrix equation of the general STM medium in Fig. 1. To solve this coupled matrix equation, we express the aperiodic/periodic spatially-variant $Z(z)$ and $Y(z)$ matrices based on the series expansion as $\textbf{W}(z)=\sum_{m=-M}^{M} \textbf{W}_m \exp[(im\varphi z)/L]$, where $\textbf{W}(z)$ represents either  $\textbf{Z}(z)$ or $\textbf{Y}(z)$, and $\textbf{Z}_m$ and $\textbf{Y}_m$ are unknown coefficients to be determined given the spatial variation of the metasurface. The dispersion relation of the unbounded general STM medium is expressed as
\begin{equation}\label{eqa:disp_rel}
 \text{det}\left\lbrace \left[ \overrightarrow{\bm{\Lambda}}_m \overrightarrow{\textbf{Z}} \overrightarrow{\bm{\Lambda}}_m \overrightarrow{\textbf{Y}}
- \textbf{I} \right] \right\rbrace   =0. 
\end{equation}

\subsection{Scattered Electromagnetic Fields}\label{sec:BCs}
Considering the TM$_{xz}$ or $E_y$ incident field in~\eqref{eqa:Ei1}, the incident magnetic field reads~\cite{Taravati_Kishk_TAP_2019}
\begin{equation}
	\mathbf{H}_\text{I}(x,z,t)=  \left[-\mathbf{\hat{x}} \cos(\theta_\text{i}) +\mathbf{\hat{z}} \sin(\theta_\text{i}) \right] \dfrac{E_0 }{\eta_1} e^{i\left( k_{x,\text{i}} x +\beta_0 z- \omega _0 t  \right)},
\end{equation}
\noindent where $\eta_1=\sqrt{\mu_0 \mu_\text{r}/(\epsilon_0\epsilon_r)}$. The electric and magnetic fields in the metasurface may be explicitly written using (6) as
\begin{subequations}
	\begin{equation}
	\mathbf{E}_\text{M}(x,z,t)=\mathbf{\hat{y}}\sum_{n,p }   \textbf{E}_{np} \left( A_{0p} e^{ i \beta_{np}^+ z}+B_{0p} e^{ -i \beta_{np}^- z} \right)  e^{  i \left( k_{x,\text{i}} x-\omega_n t \right)} ,
	\label{eqa:A-E_mod_field}
	\end{equation}
	and
	\begin{equation}
	\begin{split}
	\mathbf{H}_\text{M}(x,z,t)=
	\dfrac{1}{k} &\sum_{n,p } \textbf{H}_{np}   \Big(\left[ -\mathbf{\hat{x}} \beta_{np}^+ + \mathbf{\hat{z}} k_{x,\text{i}} \right]   A_{0p}  e^{ i \beta_{np}^+ z}\\
	&+ \left[ \mathbf{\hat{x}} \beta_{np}^- + \mathbf{\hat{z}} k_{x,\text{i}} \right] B_{0p}  e^{ -i \beta_{np}^- z} \Big)  e^{  i \left( k_{x,\text{i}} x-\omega_n t \right)} .
	\end{split}
	\label{eqa:A-H_mod_field}
	\end{equation}
\end{subequations}
\begin{subequations}\label{eqa:T_forward_backward}
	\begin{equation}\label{eqa:T_forward}
	A_{0p}=\frac{E_\text{0}  k_\text{1} [\cos(\theta_\text{i})+\cos(\theta_{\text{r}0}) ]/ \left[ -\eta_1 \beta_{0p}^+ +k_{1} Z_{p} \cos(\theta_{\text{r}0}) \right] }{  1 + \frac{\eta_1  \beta_{0p}^- +k_{1} Z_{p} \cos(\theta_{\text{r}0}) }{ -\eta_1 \beta_{0p}^+ +k_{1} Z_{p} \cos(\theta_{\text{r}0})  }  
		 \frac{\eta_3 \beta_{0p}^+ -k_{3} Z_{p} \cos(\theta_{\text{t}0}) }{\eta_3  \beta_{0p}^- +k_{3} Z_{p} \cos(\theta_{\text{t}0}) }e^{i (\beta_{0p}^+ + \beta_{0p}^-) L} },
	\end{equation}\begin{equation}\label{eqa:R_forward}
	B_{0p}=A_{0p} e^{i (\beta_{0p}^+ + \beta_{0p}^-) L}  \frac{\eta_3  \beta_{0p}^+ -k_{3} Z_{p} \cos(\theta_{\text{t}0}) }{\eta_3 \beta_{0p}^- +k_{3} Z_{p} \cos(\theta_{\text{t}0}) },
	\end{equation}
\end{subequations}
where $Z_{p}=E_{0p}/H_{0p}$.
\noindent for the backward problem, where $k_1=\omega_0\sqrt{\epsilon_\text{r,1}\mu_\text{r,1}}/c$ and $k_3=\omega_0\sqrt{\epsilon_\text{r,3}\mu_\text{r,3}}/c$ is the spatial frequency in the unmodulated media. From this point, the scattered fields in the unmodulated regions, are found as
\begin{subequations}\label{eqa:E_RT_forward}
	\begin{equation}\label{eqa:E_RF}
	\mathbf{E}_\text{R} = \mathbf{\hat{y}} \sum\limits_{n,p } \left[    E_{np} \left( A_{0p}+ B_{0p}  \right) - E_0 \delta _{n0}\right] e^{i \left[ k_{x,\text{i}} x- k_{z,\text{r}n} z  -\omega_n  t \right] },
	\end{equation}
	\begin{equation}\label{eqa:E_TF}
	\mathbf{E}_\text{T}=  \mathbf{\hat{y}} \sum\limits_{n,p }  E_{np} \left( A_{0p} e^{i \beta_{np} L} +B_{0p} e^{-i \beta_{np} L} \right) e^{i \left[ k_{x,\text{i}} x- k_{z,\text{t}n} z  -\omega_n  t \right] }.
	\end{equation}
\end{subequations}
where $k_{0n}=(\omega_0+n\Omega/v_\text{b})$.

\subsection{ST Decomposition}
The scattering angles of the different space-time harmonics (STHs) may be achieved from the Helmholtz relations where $k_{1}^2 \sin^2(\theta_{\text{i}})+k_{1n}^2 \cos^2(\theta_{\text{r}n})=k_{1n}^2$ and $k_{3}^2 \sin^2(\theta_{\text{i}})+k_{3n}^2 \cos^2(\theta_{\text{t}n})=k_{3n}^2$, where $\theta_{\text{r}n}$ and $\theta_{\text{t}n}$ are the reflection and transmission angles of the $n^\text{th}$ STH, yielding
\begin{equation}\label{eqa:refl_trans_angl}
\theta_{\text{r}n}=\theta_{\text{t},n}=\sin^{-1} \left(\frac{\sin(\theta_\text{i})}{1+n\Omega /\omega_0} \right) ,
\end{equation}

Equation~\eqref{eqa:refl_trans_angl} reveals the ST spectral decomposition of the scattered wave. For $k_1=k_2=k_3=k_0$, the reflection and transmission angles for a given harmonic $n$ are equal, given the equal tangential wavenumber, $k_{x,\text{i}}=k_0\sin(\theta_\text{i})$ in all the regions. It may be achieved from Eq.~\eqref{eqa:refl_trans_angl} that the harmonics in the $n$-interval $[\omega_0(\sin(\theta_\text{i})-1)/\Omega,+\infty]$ are scattered (reflected and transmitted) at angles ranging from $0$ to $\pi/2$ through $\theta_\text{i}$ for $n=0$, whereas the STHs outside of this interval represent imaginary $k_{znp}^\pm$ and are thus not scattered but rather propagate as surface waves along the boundary of the metasurface. The scattering angles of the STHs inside the modulated medium read
\begin{equation}\label{eqa:mod_angl}
\theta_{np}^\pm=\tan^{-1} \left(\frac{k_{x2}}{k_{znp}^\pm} \right)=\tan^{-1} \left(\frac{k_{02} \sin(\theta_{\text{i}})}{\beta_{0p}^\pm \pm n q }\right).
\end{equation}
\begin{figure}
\includegraphics[width=1\columnwidth]{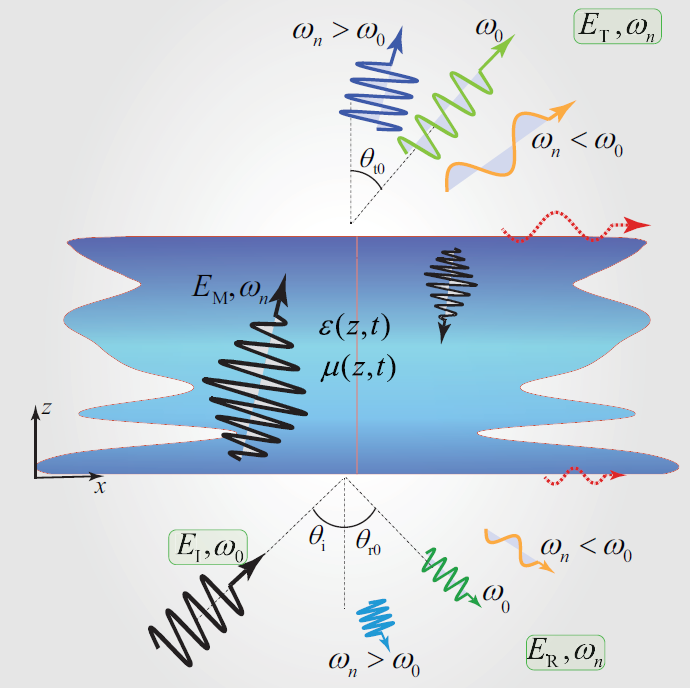}
	\caption{ST decomposition resulting from the oblique incidence to a STM metasurface~\cite{Taravati_Kishk_TAP_2019}.}
	\label{fig:decompos}
\end{figure}

\subsection{Results and Discussion}\label{sec:Res}
\begin{figure}
\includegraphics[width=1\columnwidth]{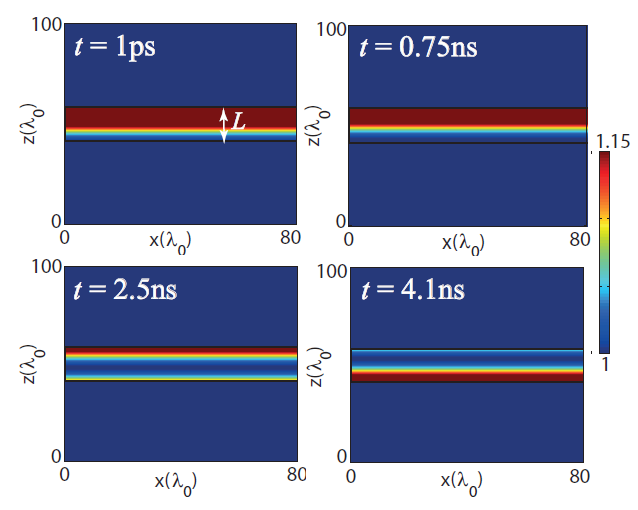}
	\caption{Spatial and temporal variation of the permittivity and permeability of a sinusoidally STM metasurface, with length $L$, sandwiched between two semi-infinite free-space regions, where $\Omega=2\pi\times0.1$~GHz, $\varGamma=1$ $\delta_{\epsilon}=\delta_{\mu}=0.15$, and for $t=1$ps, $t=0.75$ns, $t=2.5$ns and $t=4.1$ns~\cite{Taravati_Kishk_TAP_2019}.}
	\label{Fig:perm_prof}
\end{figure}

This section presents the analytical and numerical investigation of wave propagation and scattering from general STM metasurfaces. As a particular case, which is easier to realize~\cite{Taravati_PRB_2017,Taravati_LWA_2017}, we consider a sinusoidally STM medium, as
\begin{subequations}\label{eqa:ap_constit}
	\begin{equation}\label{eqa:permit}
	\epsilon(z,t)=\epsilon_0 \epsilon_\text{r} \left[1+\delta_{\epsilon} \sin(q z-\Omega t) \right],
	\end{equation}
	\begin{equation}\label{eqa:permeab}
	\mu(z,t)=\mu_0 \mu_\text{r}\left[1+\delta_{\mu} \sin(q z-\Omega t) \right],
	\end{equation}
\end{subequations}
where $\delta_{\epsilon}$ and $\delta_{\mu}$ represent respectively the permittivity and permeability modulation strengths. Figure~\ref{Fig:perm_prof} plots the spatial and temporal variation of the permittivity and permeability of a sinusoidally STM metasurface, with length of $L$, sandwiched between two semi-infinite free-space regions. The metasurface assumes $\Omega=2\pi\times0.1$~GHz, $\varGamma=1$, $\delta_{\epsilon}=\delta_{\mu}=0.15$, $t=1$ps, $t=0.75$ns, $t=2.5$ns and $t=4.1$ns. Such a metasurface is characterized with a ST-varying intrinsic impedance~\cite{Taravati_PRAp_2018}, i.e.,
\begin{equation}\label{eqa:imp}
\eta(z,t)=\sqrt{\dfrac{\mu_0 \mu_\text{r}\left[1+\delta_{\mu} \sin(q z-\Omega t) \right]}{\epsilon_0 \epsilon_\text{r} \left[1+\delta_{\epsilon} \sin(q z-\Omega t) \right]}}\Bigg|_{\delta_{\mu}=\delta_{\epsilon}}=\eta_0 \eta_\text{r}
\end{equation}
Equation~\eqref{eqa:imp} reveals that such a STM metasurface exhibits zero space- and time local reflections as the intrinsic impedance of the metasurface is ST-independent.

\subsection{Dispersion Diagrams for a Sinusoidal STM Medium}\label{sec:disp}
\begin{figure}
\includegraphics[width=1\columnwidth]{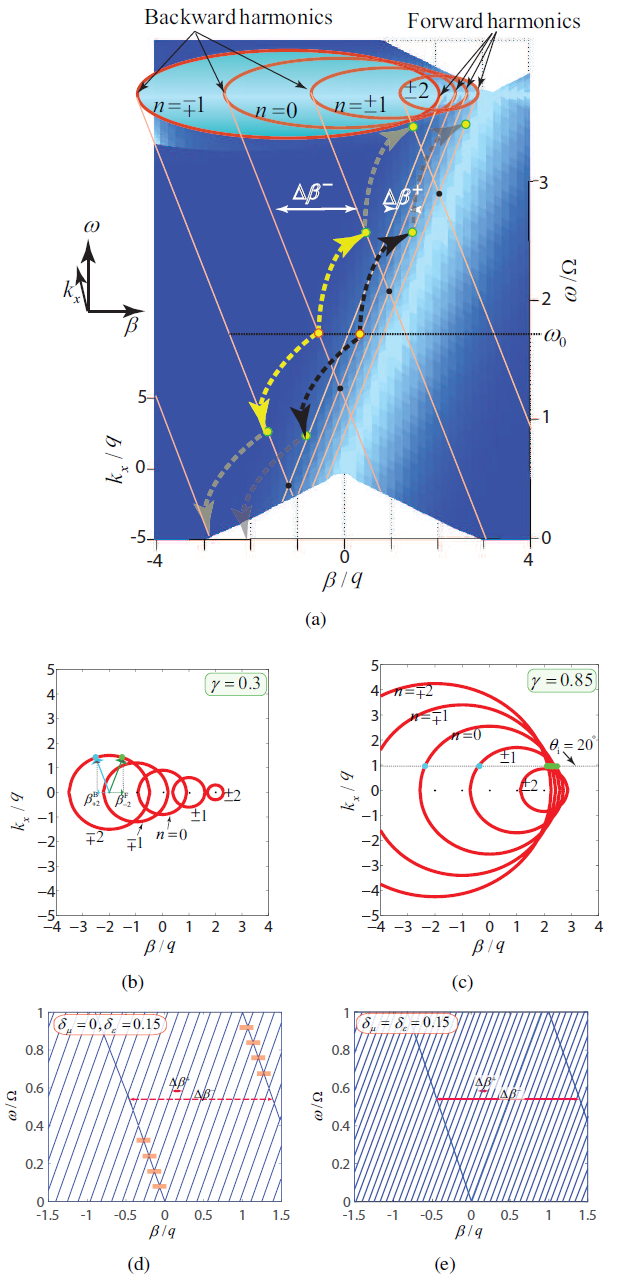}
	\caption{Analytical dispersion diagram of the sinusoidally STM medium with the electric permittivity and magnetic permeability in~\eqref{eqa:ap_constit} in the quasisonic regime, i.e., $\varGamma=0.85$. (a)~Three dimensional dispersion diagram constituted of an array of double-cones for $\delta_{\epsilon}=\delta_{\mu}\rightarrow 0$~\cite{Taravati_Kishk_TAP_2019}. (b)~and (c)~Isofrequency diagrams at $\omega/\Omega=3$ and for $\delta_{\epsilon}=\delta_{\mu}\rightarrow 0$, respectively, for the subsonic regime ($\varGamma=0.3$) and the quasisonic regime ($\varGamma=0.85$)~\cite{Taravati_Kishk_TAP_2019}.
	(d)~Normal incidence ($k_{x,\text{i}}=0$) dispersion diagram of the conventional ST permittivity-modulated metasurface, i.e., $\delta_{\epsilon}=0.15$ and $\delta_{\mu}=0$. This diagram is constituted of an infinite set of periodic forward lines with the distance $\Delta \beta^+$ and an infinite set of periodic backward lines with the distance  $\Delta \beta^-$, and periodic electromagnetic band gaps appear at the intersection of the forward and backward STHs~\cite{Taravati_Kishk_TAP_2019}. (e)~Normal incidence ($k_{x,\text{i}}=0$) dispersion diagram of the equilibrated STM metasurface with $\delta_{\mu}=\delta_{\epsilon}=0.15$, yielding enhanced nonreciprocity, $\Delta \beta^-/\Delta \beta^+$, and zero electromagnetic band gaps~\cite{Taravati_Kishk_TAP_2019}.}
	\label{fig:10}
\end{figure}

Wave propagation in an unbounded STM medium may be best investigated by the analysis of its dispersion diagrams. Figure~\ref{fig:10}(a) plots the three dimensional dispersion diagram using~\eqref{eqa:disp_rel} for sinusoidally STM medium with the electric permittivity and magnetic permeability in~\eqref{eqa:ap_constit} for $\delta_{\mu}=\delta_{\epsilon}=0.15$ and $\varGamma=0.85$, exhibiting $\beta_0(\omega_0,k_{x,\text{i}})$. This diagram is constituted of an array of cones with different diameters and different origins. For a fixed frequency $\omega_0$, this 3D diagram gives a 2D diagram constituted of an array of circles, each of which representing a STH (shown in Figs.~\ref{fig:10}(b) and~\ref{fig:10}(b)). However, for a fixed $k_{x,\text{i}}$, the 3D diagram in Fig.~\ref{fig:10}(a) provides an infinite set of forward lines ($\partial \beta/\partial \omega>0$) with the distance of $\Delta \beta^+/q=1-\varGamma$, and an infinite set of backward lines ($\partial \beta/\partial \omega<0$) with the distance of $\Delta \beta^-/q=1+\varGamma$(shown in Figs.~\ref{fig:10}(d) and~\ref{fig:10}(e). The nonreciprocity of the medium is proportional to the ratio $\Delta \beta^-/\Delta \beta^+$. In the sonic regime, where $\varGamma \rightarrow 1$, the distance between the forward lines tends to zero $\Delta \beta^+/q=1-\varGamma\rightarrow 0$, yielding a \textit{strong} exchange of energy and momentum between the forward harmonics, whereas the distance between the backward lines $\Delta \beta^-/q=1+\varGamma\rightarrow 2$, yielding a \textit{weak} exchange of energy and momentum between the backward harmonics. As a consequence, exciting the medium at the temporal and spatial frequencies [$\omega_0$,$\beta_0$] results in a strong cascade transition of energy and momentum to higher order forward harmonics [$\omega_0+n\Omega$,$|\beta_0|+nq$] (shown with black dashed arrows in Fig.~\ref{fig:10}(a)), and a weak cascade transition of energy and momentum to higher order backward harmonics [$\omega_0+n\Omega$,$-|\beta_0|+nq$] (shown with yellow dashed arrows in Fig.~\ref{fig:10}(a)), with $n$ being any integer. Hence, in the limiting case of the sonic regime where the strongest nonreciprocity ($\Delta \beta^-/\Delta \beta^+$) is provided by the STM medium, there still exist a weak transition of backward STHs as the distance between backward harmonics $\Delta \beta^-/q$ does not acquire an infinite value, rather tends to $\Delta \beta^-/q\rightarrow 2$.

For a nonzero permittivity (or permeability) modulation strength (Fig.~\ref{fig:10}(d)), i.e., $\delta_{\mu}\neq\delta_{\epsilon}$, the equilibrium between the electric permittivity and magnetic permeability of the medium will be lost. Hence, asymmetric (with respect to $\beta/q=0$) and unequal electromagnetic band gaps open up at the synchronization points between the forward and backward harmonics~\cite{Taravati_PRAp_2018}. These band gaps correspond to frequencies where the incident wave will be largely reflected by the medium. However, as we equally increase both the permittivity and permeability modulation strength (Fig.~\ref{fig:10}(e)), equilibrium in the electromagnetic properties of the medium occurs and the electromagnetic band gaps disappear.

\subsection{FDTD Numerical Simulation}\label{sec:num}
\begin{figure}
	\begin{center}
		\includegraphics[width=1\columnwidth]{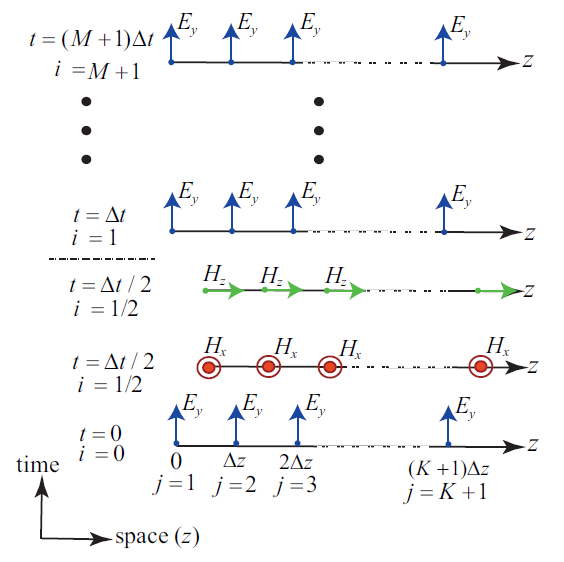} 
		\caption{General representation of the finite-difference time-domain scheme for numerical simulation of the oblique incidence of an $E_y$ wave to STM slab~\cite{Taravati_Kishk_PRB_2018}.} 
		\label{Fig:FDTD}
	\end{center}
\end{figure}

We next verify the above theory by FDTD numerical simulation of the dynamic process through solving Maxwell's equations. Figure~\ref{Fig:FDTD} plots the implemented finite-difference time-domain scheme for numerical simulation of the oblique wave impinging to the STM slab. We first discretize the medium to $K+1$ spatial samples and $M+1$ temporal samples, with the steps of $\Delta z$ and $\Delta t$, respectively.
Next, the finite-difference discretized form of the first two Maxwell's equations for the electric and magnetic fields in~\eqref{eqa:Max1} and~\eqref{eqa:Max2} are simplified to
\begin{subequations}
	\begin{equation}\label{eqa:num_Max1c}
	\begin{split}
	H_x\lvert_{j+1/2}^{i+1/2}=&\left(1-\Delta t  \right)	H_x\lvert_{j+1/2}^{i-1/2}+\dfrac{\Delta t}{ \mu_0 \Delta z} \left( E_y\lvert_{j+1}^{i}-E_y\lvert_{j}^{i} \right) 
	\end{split}	
	\end{equation}
	\begin{equation}\label{eqa:num_Max1d}
	\begin{split}
	H_z\lvert_{j+1/2}^{i+1/2}=&\left(1-\Delta t  \right)	H_z\lvert_{j+1/2}^{i-1/2}-\dfrac{\Delta t}{ \mu_0 \Delta z} \left( E_y\lvert_{j+1}^{i}-E_y\lvert_{j}^{i} \right) 
	\end{split}	
	\end{equation}
	\begin{equation}\label{eqa:num_Max2c}
	\begin{split}
	E_y& \lvert_{j}^{i+1}= \left(1- \dfrac{\Delta t\epsilon'\lvert_{j}^{i}}{\epsilon \lvert_{j}^{i+1/2}}  \right) E_y \lvert_{j}^{i}+ \dfrac{\Delta t/\Delta z}{ \epsilon \lvert_{j}^{i+1/2}} \\
	&.\left[ \left( H_x \lvert_{j+1/2}^{i+1/2}-H_x\lvert_{j-1/2}^{i+1/2} \right) -\left( H_z \lvert_{j+1/2}^{i+1/2}-H_z\lvert_{j-1/2}^{i+1/2} \right)\right] 
	\end{split}	
	\end{equation}
\end{subequations}
\noindent where $\epsilon'=\partial \epsilon(z,t)/\partial t=-\Omega \delta_{\epsilon} \cos(qz-\Omega t)$.

\subsection{Advanced Wave Engineering Based on Unidirectional Frequency Generation and ST Decomposition}
This section investigates the wave transmission and reflection from STM media using the FDTD numerical simulation. We consider oblique incidence to general STM metasurfaces, and compare the numerical results with the analytical solution provided in Sec.~\ref{sec:anal_sol}. A plane wave with temporal frequency $\omega_0=2\pi\times3$~GHz is propagating along the $+z$-direction under an angle of incidence of $\theta_\text{i}=25^\circ$, and impinges to the conventional ST permittivity-modulated metasurface in Fig.~\ref{Fig:perm_prof} with the constitutive parameters in~\eqref{eqa:ap_constit}, where $\Omega=2\pi\times0.1$~GHz, $\varGamma=1$, $L=16 \lambda_0$.

\subsubsection{Periodic ST Modulation}
\begin{figure}
\includegraphics[width=1\columnwidth]{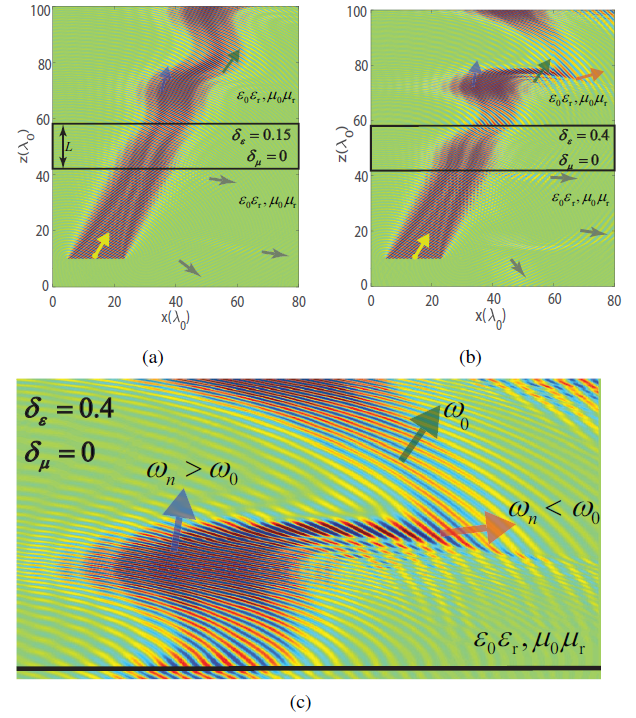}  
	\caption{Electric field distribution, $E_y$, for the forward oblique excitation of a plane wave with frequency $\omega_0=2\pi\times3$~GHz to ST permittivity-modulated metasurface ($\delta_{\mu}=0$) in Fig.~\ref{Fig:perm_prof} with $\Omega=2\pi\times0.1$~GHz, $\varGamma=1$, $L=16 \lambda$, $\theta_\text{i}=25^\circ$ and at $t=200$~ns. (a)~Moderate modulation strength $\delta_{\epsilon}=0.15$~\cite{Taravati_Kishk_TAP_2019}. (b)~Large modulation strength $\delta_{\epsilon}=0.4$~\cite{Taravati_Kishk_TAP_2019}. (c)~A magnified image highlighting the spatiotemporally decomposed transmitted STHs in (b)~\cite{Taravati_Kishk_TAP_2019}.}
	\label{Fig:12}
\end{figure}

Figure~\ref{Fig:12}(a) plots the electric field distribution for the scattered field from the ST permittivity-modulated metasurface with $\delta_{\epsilon}=0.15$ and $\delta_{\mu}=0$. The transmitted field from the metasurface, on top of this figure, is constituted of a set of STHs, i.e. $\omega_n=2\pi\times(3+0.1n)$~GHz, with $n$ being any integer. Small reflections are seen at the bottom of this figure, highlighted with arrows, which are due to the local space and time reflections inside the STM metasurface. In addition, it is however obvious that the frequency generation and decomposition is not strong. To achieve stronger frequency generation and ST decomposition, we may increase the modulation strength or the metasurface length. 

Figures~\ref{Fig:12}(b) and~\ref{Fig:12}(c) plot the result for the same metasurface as Fig.~\ref{Fig:12}(a) except for a higher modulation strength of $\delta_{\epsilon}=0.4$, with $\delta_{\mu}=0$. ST decomposition of the transmitted wave is pronounced in this figure, where high frequency harmonics ($\omega_n>\omega_0$) are transmitted under the angle of transmission of $\theta_\text{t,$n$}<\omega_n$, and low frequency harmonics ($\omega_n<\omega_0$) are transmitted under the angle of transmission of $\theta_\text{t,$n$}>\omega_n$. As expected, for such a metasurface with strong permittivity modulation, $\delta_{\epsilon}=0.4$, the reflected STHs at the bottom of the figure, are more pronounced.

\begin{figure*}
\includegraphics[width=2\columnwidth]{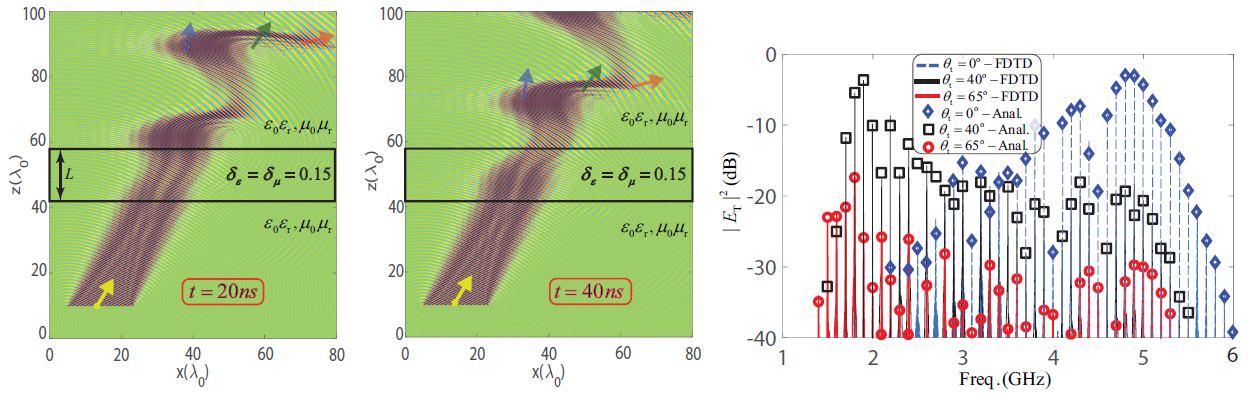}
\caption{Forward oblique excitation of the equilibrated STM metasurface in Fig.~\ref{Fig:perm_prof}, with a plane wave of frequency $\omega_0=2\pi\times3$~GHz, with $\delta_{\mu}=\delta_{\epsilon}=0.15$ and for $\theta_\text{i}=25^\circ$. (a)~Numerical simulation result for the electric field distribution, $E_y$, at $t=20$~ns~\cite{Taravati_Kishk_TAP_2019}. (b)~Numerical simulation result for the electric field distribution, $E_y$, at $t=40$ ns~\cite{Taravati_Kishk_TAP_2019}. (c)~Comparison of the analytical and numerical solutions for the spectrum of the transmitted electric field, at $t=200$~ns, for three different angles, i.e. $\theta_\text{t}=0,40,65^\circ$ for $\varGamma=0.85$~\cite{Taravati_Kishk_TAP_2019}.}
	\label{Fig:13}
\end{figure*}

Next, we investigate the field scattering from an equilibrated STM metasurface, i.e. $\delta_{\mu}=\delta_{\epsilon}=0.15$. Figure~\ref{Fig:13}(a) plots the electric field distribution inside and scattered from this metasurface at $t=20$~ns. It may be seen from this figure that, an equilibrated STM metasurface with $\delta_{\mu}=\delta_{\epsilon}=0.15$ exhibits similar frequency generation and ST decomposition as a conventional ST permittivity-modulated metasurface with much stronger modulation strength of $\delta_{\epsilon}=0.4$ (shown in Fig.~\ref{Fig:12}(b)). It may be shown that an equilibrated STM metasurface with $\delta_{\mu}=\delta_{\epsilon}=0.15$ may be realized with the same amount of pumping energy required for the realization of conventional permittivity-modulated metasurface with $\delta_{\epsilon}=0.15$ and $\delta_{\mu}=0$~\cite{Taravati_PRAp_2018}. Moreover, as we see in Fig.~\ref{Fig:13}(a), the equilibrated metasurface exhibits zero local space and time reflection. The strict zero reflection from such a metasurface can be shown by increasing the number of space and time samples in the numerical scheme.
\begin{figure*}
\includegraphics[width=2\columnwidth]{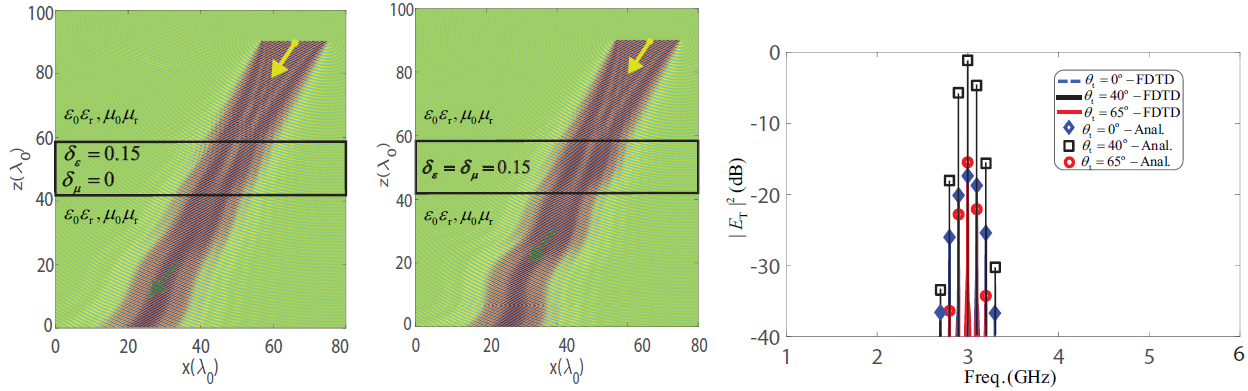}
	\caption{Backward oblique excitation of the metasurface in Fig.~\ref{Fig:13}, with a plane wave of frequency $\omega_0=2\pi\times3$~GHz, with $\delta_{\epsilon}=0.15$, $\theta_\text{i}=25^\circ$ and at $t=200$~ns. (a)~$\delta_{\mu}=0$~\cite{Taravati_Kishk_TAP_2019}. (b)~$\delta_{\mu}=\delta_{\epsilon}=0.15$~\cite{Taravati_Kishk_TAP_2019}. (c)~Analytical and numerical solutions for the spectrum of the transmitted electric field at $t=200$~ns at three different angles, i.e. $\theta_\text{t}=0,40,65^\circ$  for $\varGamma=0.85$~\cite{Taravati_Kishk_TAP_2019}.}
	\label{Fig:14}
\end{figure*}

Figure~\ref{Fig:13}(b) shows the numerical simulation result for the electric field distribution inside the STM metasurface and scattered from the same metasurface as in Fig.~\ref{Fig:13}(a) except at $t=40$~ns. As expected, such a time-varying metasurface presents a time-varying beam for the transmitted field, as well as the reflected field.

Figure~\ref{Fig:13}(c) compares the analytical and numerical solutions for the spectrum of the transmitted electric field for forward incidence to the metasurface, at $t=200$~ns, for three different angles, i.e. $\theta_\text{t}=0,40,65^\circ$. Interestingly, the energy of the incident wave at $\omega_0=2\pi\times3$~GHz is strongly transitioned to higher STHs for $\theta_\text{t}=0^\circ$, and to lower STHs for $\theta_\text{t}=40^\circ$ and $\theta_\text{t}=65^\circ$. This figure reveals that the STM metasurface may be cast as a subharmonic mixer.

Figures~\ref{Fig:14}(a) and~\ref{Fig:14}(b) respectively plot the numerical simulation result for the electric wave scattering from the conventional (same metasurface as in Fig.~\ref{Fig:12}) and equilibrated (same metasurface as in Fig.~\ref{Fig:13}) for the backward excitation. It may be seen that for both cases, the frequency generation and ST decomposition of the transmitted field are negligible. As a result, such STM metasurface, provides nonreciprocal frequency generation and ST decomposition.

Figure~\ref{Fig:14}(c) compares the analytical and numerical solutions for the spectrum of the transmitted electric field for backward incidence to the equilibrated metasurface (same metasurface as in Figs.~\ref{Fig:13} and~\ref{Fig:14}(a) and~\ref{Fig:14}(b) for three different angles, i.e. $\theta_\text{t}=0,40,65^\circ$. Figure~\ref{Fig:14}(c) shows that, the energy of the incident wave at $\omega_0=2\pi\times3$~GHz is weakly transited to higher and lower STHs.

\subsubsection{Controlling the STHs Using Aperiodic ST Modulation}\label{sec:aper}
\begin{figure*}
\includegraphics[width=2\columnwidth]{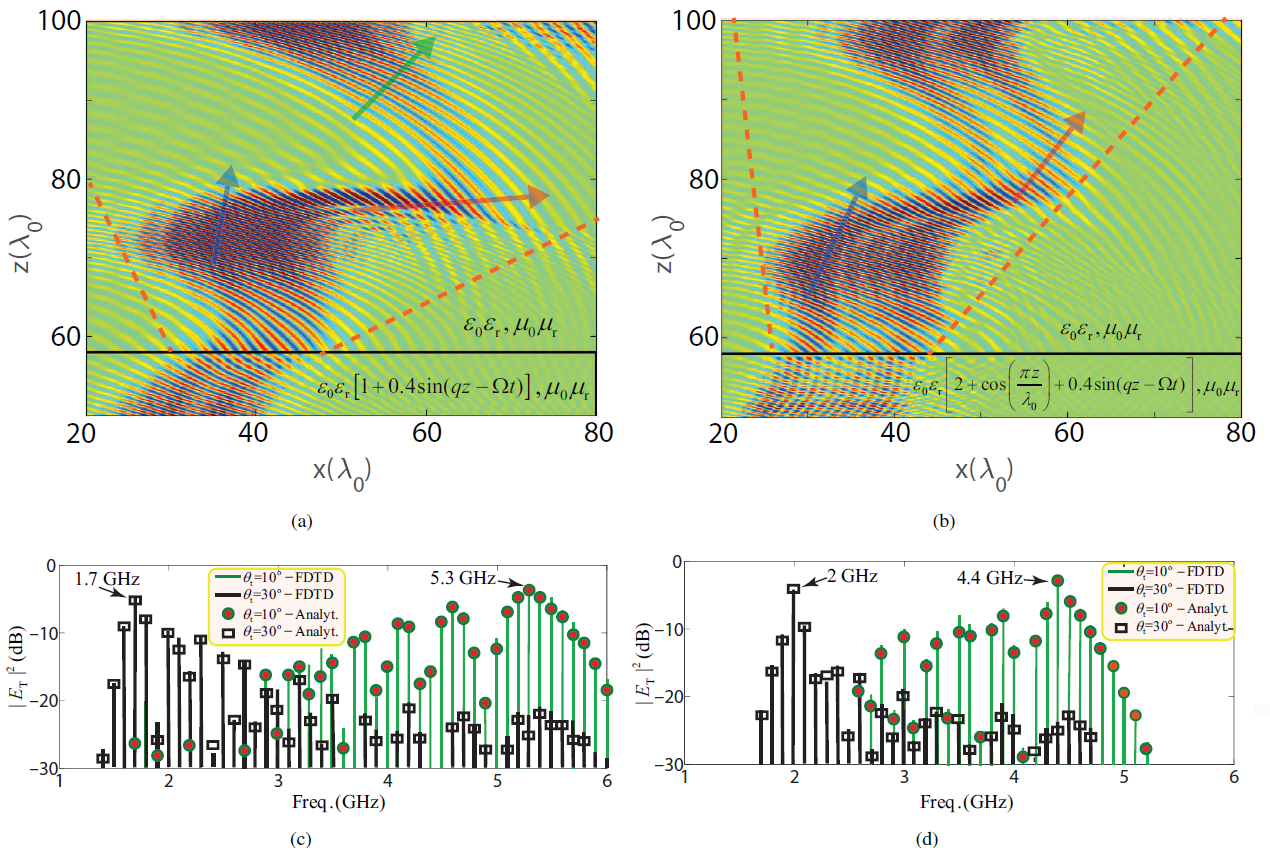} 
	\caption{Numerical simulation results for the electric field distribution, $E_y$, for the forward oblique excitation under the angle of incidence $\theta_\text{i}=25^\circ$, $\omega_0=2\pi\times3$~GHz, where $\delta_{\epsilon}=0.4$, $\delta_{\mu}=0$, $\Omega=2\pi\times0.1$~GHz, $\varGamma=0.85$, $L=16 \lambda$, and at $t=200$~ns. (a)~and (c)~Time domain and frequency domain results, respectively, for the periodic ST permittivity-modulated medium~\cite{Taravati_Kishk_TAP_2019}. (b)~and (d)~ Time domain and frequency domain results, respectively, for the quasi-aperiodic STM medium~\cite{Taravati_Kishk_TAP_2019}.}
	\label{Fig:15}
\end{figure*}

To best control the transmitted STHs, one may use an aperiodic ST modulation~\cite{Taravati_PRB_Mixer_2018}. The aperiodic ST modulation provides the leverage for the suppression of undesired STHs~\cite{Taravati_PRB_Mixer_2018}, as well as the capability to change the angles and amplitudes of the transmitted STHs. An appropriate periodic/aperiodic spatial variation added to the periodic ST modulation in the previous section may provide the sufficient design flexibility to control the properties of the transmitted STHs. Such an aperiodic spatial variation may be specified through a similar synthesis procedure reported in~\cite{Taravati_PRB_Mixer_2018}. In order to show the effect of a nonuniform (periodic/aperiodic) spatial variation on the transmitted STHs, we consider a particular case of a quasi-aperiodic STM permittivity 
\begin{equation}\label{eqa:permit_a}
	\epsilon(z,t)=\epsilon_0 \epsilon_\text{r} \left[2+M\cos(\sigma \frac{2\pi}{\lambda_0} z) +\delta_{\epsilon} \sin(q z-\Omega t) \right],
\end{equation}
and constant permeability $\mu(z,t)=\mu_0 \mu_\text{r}$, where $\lambda_0=v_\text{b}/\omega_0$. We aim to achieve a desired transmission, in which $\omega_{-10}$ and $\omega_{14}$ are respectively transmitted under the transmission angles of $\theta_{\text{T},-10}=30^\circ$ and $\theta_{\text{T},14}=10^\circ$, where $\omega_{n}>\omega_{20}$ and $\omega_{n}<\omega_{-10}$ are attenuated. We first look for a periodic spatial profile. Following the synthesis procedure in~\cite{Taravati_PRB_Mixer_2018}, $M=1$ and $\sigma$ provide a suitable response. 
\indent Figures~\ref{Fig:15}(a) and~\ref{Fig:15}(b) show the numerical results for the time domain response for the transmission of the STHs, respectively, through the periodic STM medium and quasi-aperiodic STM. It may be observed that the quasi-aperiodic STM medium has provided quite different transmitted STHs than the periodic case, i.e., with different amplitudes and different angles of transmissions. This effect may be best seen in the frequency domain. Figures~\ref{Fig:15}(c) and~\ref{Fig:15}(d) show the analytical and numerical simulation results for the frequency domain transmitted STHs, respectively, through the periodic STM medium and quasi-aperiodic STM. As specified, the strongest harmonic at $\theta_{\text{T},14}=10^\circ$ is at $\omega_{14}=4.4$ GHz, where the strongest harmonic at $\theta_{\text{T},-10}=10^\circ$ is at $\omega_{14}=2$ GHz, and $\omega_{n}>\omega_{20}$ and $\omega_{n}<\omega_{-10}$ are significantly attenuated. This demonstrates the capability of the obliquely illuminated STM in wave engineering. It should be noted that, an aperiodic profile may provide better response in terms of the suppression of undesired STHs, but here we consider the simplest case which is easier to design and implement.

\section{ST Diffraction Metagratings}
Light diffraction by spatially periodic structures is a fundamental phenomenon in optics and is of great importance in a variety of engineering applications~\cite{gaylord1982planar}. Here, we first introduce the concept of generalized periodic gratings~\cite{taravati_PRApp_2019}. Such gratings are varying periodically in both space and time, representing the generalized version of standard conventional static (time-invariant) spatially varying gratings. Different from the ST metasurfaces in Secs.~\ref{sec:Res} where the ST modulation is applied along the $z$ (perpendicular) direction, here the STP grating is periodic along the $x$ (transverse). The analytical solution in Sec.~\ref{sec:anal_sol} is for general aperiodic ST metasurfaces where the metasurfaces lie outside the diffraction regime. As a result, here we present a deep analysis on the functionality of STP gratings in the diffraction regime based on the modal analysis for electromagnetic waves inside a STP grating along with the wavevector-diagram analysis for diffracted waves outside the STP grating.
\begin{figure}
	\begin{center}
\includegraphics[width=1\columnwidth]{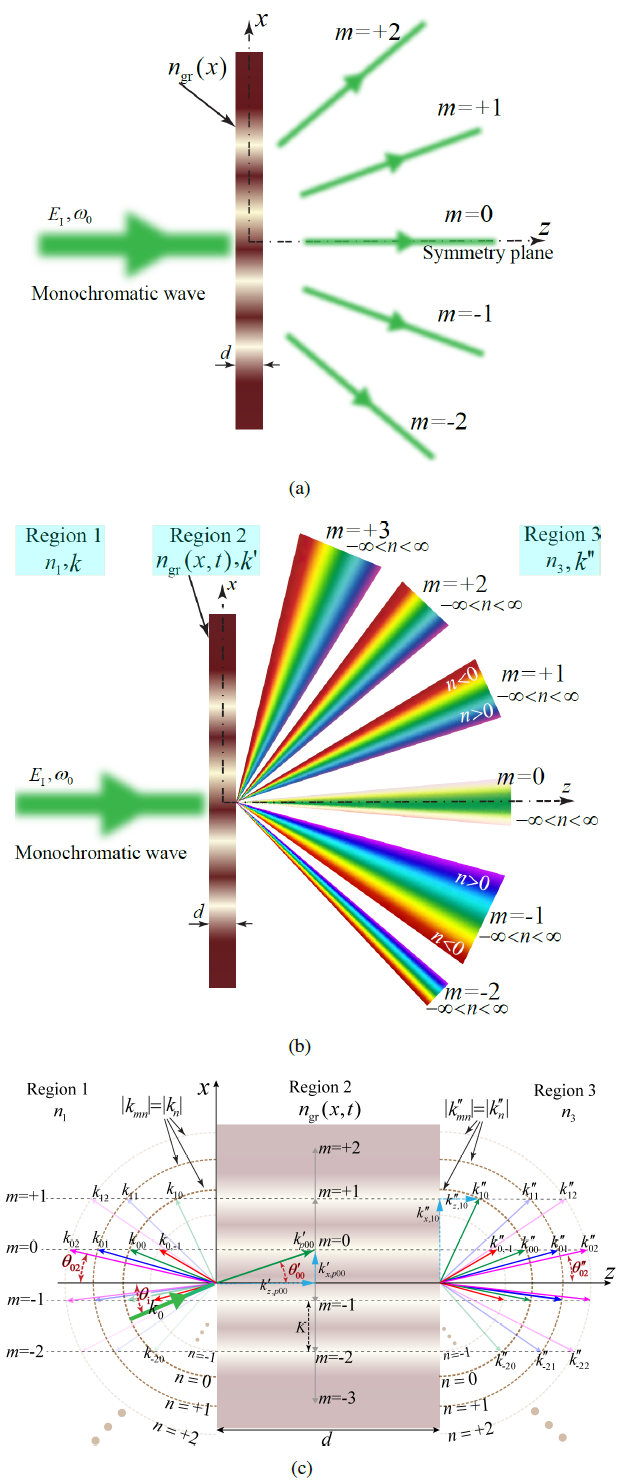}
		\caption{Diffraction from a transmissive grating for a monochromatic incident wave. (a) Conventional spatial diffraction grating with $n_\text{gr}^2(x)= \epsilon_\text{gr}(x)=f_\text{per} (x)$~\cite{taravati_PRApp_2019}.
			(b) Generalized STP diffraction grating with $n_\text{gr}^2(x,t)= \epsilon_\text{gr}(x,t)=f\left(f_\text{1,per} (x),f_\text{2,per} (t) \right)$~\cite{taravati_PRApp_2019}. (c) Wavevector isofrequency diagram for the diffraction from a STP diffraction grating~\cite{taravati_PRApp_2019}.}
		\label{Fig:16}
	\end{center}
\end{figure}

Figure~\ref{Fig:16}(a) depicts the wave diffraction from a conventional transmissive planar spatially periodic diffraction gratings. The conventional static grating in Fig.~\ref{Fig:16}(a) possesses a relative electric permittivity in the region from $z = 0$ to $z = d$ given by $n_\text{gr}^2(x)= \epsilon_\text{gr}(x)=f_\text{per} (x)$, where $f_\text{per} (x)$ is a periodic function of $x$, e.g., a sinusoidal, binary (square), or sawtooth function. Electromagnetic waves always travel in straight lines, but when passing near an obstruction they tend to bend around that obstruction and spread out. The diffraction phenomenon occurs when an electromagnetic wave passes by a corner or through a slit or grating that has an optical size comparable to the wavelength. The diffraction by a grating is a specialized case of wave scattering, where an object with regularly repeating features yields an orderly diffraction of the electromagnetic wave in a pattern consisting of a set of diffraction orders $m$.

As shown in Fig.~\ref{Fig:16}(a), considering normal incidence of the input wave ($\theta_\text{i}=0$), a symmetric diffraction pattern with respect to $x=0$ will be produced by conventional static gratings, possessing a symmetric profile with respect to the $x=0$ axis. An asymmetric diffraction pattern for normal incidence can be achieved by an asymmetric static periodic metagratings~\cite{popov2018controlling,popov2019constructing}. However, gratings with symmetric and asymmetric profiles are both restricted by the Lortentz reciprocity theorem, and therefore, possess reciprocal diffraction transmission response. The symemtry of the diffraction pattern in conventional periodic static gratings includes the symmetry in both the angles of diffraction orders $\theta_{m}$ (e.g., $\theta_{+2}=\theta_{-2}$) and the symmetry in the intensity of the diffracted orders $P_m$  (e.g., $P_{+2}=P_{-2}$). In addition, assuming a monochromatic input wave with temporal frequency $\omega_0$, no change in the temporal frequency of the incident field occurs, and hence, the diffracted orders share the same temporal frequency of $\omega_0$.

Now, consider the transmissive planar STP diffraction grating shown in Fig.~\ref{Fig:16}(b). This figure shows a generic representation of the ST diffraction from a STP diffraction grating, which is distinctly different from the spatial diffraction from a conventional space-periodic diffraction grating in Fig.~\ref{Fig:16}(a). The grating is interfaced with two semi-infinite dielectric regions, i.e., region 1 is characterized with the refractive index $n_1$ and wavenumber $k$, and region 3 characterized with the refractive index $n_3$ and wavenumber $k''$. The relative electric permittivity of this STP grating is periodic in both space and time, with temporal frequency $\Omega$ and spatial frequency $K$, given by $n_\text{gr}^2(x,t)= \epsilon_\text{gr}(x,t)=f\left(f_\text{1,per} (x),f_\text{2,per} (t) \right)$, where $f_\text{1,per} (x)$ and $f_\text{2,per} (t)$ are periodic functions of space (in the $x$ direction) and time, respectively. The wavenumber in region 2 (inside the STP grating) is denoted by $k'$.

Assuming normal (or oblique) incidence of the input wave, the STP gratings (shown in Fig.~\ref{Fig:16}(b)) produces an asymmetric diffraction pattern with respect to $x=0$. This asymmetry in the diffraction pattern is due to the asymmetric ST profile of the structure provided by the ST modulation. The asymmetry of the diffraction pattern extends to both the diffraction angles of diffracted orders $\theta_{m}$ (e.g., $\theta_{m=+2} \neq \theta_{m-2}$) and the intensities of the diffracted orders $P_m$  (e.g., $P_{m=+2} \neq P_{m=-2}$). Furthermore, the time-variation of the grating (with frequency $\Omega$) results in the generation of new frequencies. Hence, assuming a monochromatic input wave with temporal frequency $\omega_0$, an infinite set of temporal frequencies will be generated inside the grating and will be diffracted, so that each spatial diffracted order ($m$) is composed of an infinite number of temporal diffraction orders $n$. As a result, for such a generalized STP diffraction grating, the diffraction characteristics are defined for each ST diffracted order ($mn$) so that the diffracted order ($mn$) is transmitted at a specified angle $\theta_{mn}$ attributed to the electric field $E_{mn}^\text{T}$.

Figure~\ref{Fig:16}(c) shows a generic illustration example of a wavevector isofrequency diagram for the diffraction from a STP diffraction grating. The grating is characterized with the spatial frequency $K$ (the spatial periodicity of the STP grating reads $\Lambda=2\pi/K$) and the temporal frequency $\Omega$. Figure~\ref{Fig:16}(c) sketches the phase matching of ST harmonic components of the total field inside the grating with propagating backward diffracted orders in region 1, and
forward diffracted orders in region 3. We assume the grating is interfaced with two semifinite dielectrics, i.e., $z\rightarrow -\infty<\text{region 1}<z=0$ and $d<\text{region 3}<z \rightarrow \infty$, respectively. Region 1, region 2 (inside the STP grating) and region 3 are, respectively, characterized with the phase velocities $v_\text{r}=c/n_1$, $v'_\text{r}=c/n_\text{av}$ and $v''_\text{r}=c/n_3$, and the wavevectors $\text{\textbf{k}}_{mn}= k_{x,mn} \mathbf{\hat{x}}+k_{z,mn} \mathbf{\hat{z}}$, $\text{\textbf{k}}'_{pmn}= k'_{x,pmn} \mathbf{\hat{x}}+k'_{z,pmn} \mathbf{\hat{z}}$, and $\text{\textbf{k}}''_{mn}= k''_{x,mn} \mathbf{\hat{x}}+k''_{z,mn} \mathbf{\hat{z}}$. Here, $c$ represents the velocity of the light in vacuum, $m$ and $n$ denote the number of the space and time harmonics, respectively, while $p$ represents the number of the mode in region 2, inside the grating (these modes only exist inside the grating). 	

The STP grating assumes oblique incidence of the $y$-polarized electric field in Eq.~\eqref{eqa:Ei1}. The $x$ component of the wavevector outside the STP grating, in region 3, reads $k''_{x,mn}=k''_n \sin(\theta''_{mn})$, where $k''_n =k''_0+n \Omega/v''_\text{r}$ and where $k''_0=\omega_0/v''_\text{r}$. The corresponding $z$ component of the wavevector in region 3 is calculated using the Helmholtz relation, as
$k''_{z,mn}=\sqrt{(k''_{mn})^2-(k''_{x,mn})^2}=k''_n \cos(\theta''_{mn})$.  The ST diffraction process may be simply interpreted as follows. The incident wave is refracted into the grating medium at $z = 0$, while generating an infinite set of time harmonics inside the grating, with frequencies $\omega_n$ corresponding to the wavevectors $k'_n=k'_0+n \Omega/v'_\text{r}$. The refracted ST plane waves in the grating are diffracted into an infinite set of plane waves traveling toward the $z = d$ boundary. The ST harmonic waves inside the grating are phase matched into propagating and evanescent waves in region 3, i.e., the $x$ components of the wavevectors of the $m$th mode in regions 1 and 3 and the $x$ component of the wavevector of the $m$th ST harmonic field in region 2 must be the same.

To determine the spatial and temporal frequencies of the diffracted orders, we consider the momentum conservation law, i.e.,
$k_{x,\text{diff}}=k''_{x,mn}=k_{x,\text{i}}+ m K$ and the energy conservation law, i.e., $\omega_\text{diff}=\omega_0+ n \Omega$, where $k_{x,\text{diff}}$ and $k_{x,\text{i}}$ denote the $x$ components of the wavevector of the diffracted and incident fields, respectively, and $\omega_\text{diff}$ and $\omega_0$ represent the temporal frequencies of the diffracted and incident fields, respectively. Then,
\begin{equation}
\left(k''_0+n \frac{\Omega}{v''_\text{r}} \right) \sin\left(\theta''_{mn} \right)=k_0 \sin(\theta_\text{i})+mK,
\end{equation} 
where $k_0=n_1\omega_0/c$. Considering $n_1=n_3$, the angle of diffraction for the forward ST diffracted orders in region 3 and the backward ST diffracted orders in region 1, i.e., the $m$th spatial and $n$th temporal harmonic, yields	
\begin{equation}\label{eqa:refl_trans_angl_1}
\sin\left(\theta''_{mn} \right)= \frac{\sin(\theta_\text{i})+m K/k_0}{1+n \Omega/\omega_0}   ,
\end{equation}

\subsection{Propagating and evanescent Orders}
For a given set of incident angles, spatial and temporal frequencies of the grating, and the wavelength of the incident beam, the grating equation may be satisfied for more than one value of $m$ and $n$. However, there exists a solution only when $|\sin\left(\theta_{mn} \right)|<1$. Diffraction orders corresponding to $m$ and $n$ satisfying 
this condition are called \textit{propagating} orders. The other orders yielding $|\sin\left(\theta_{mn} \right)|>1$ correspond to imaginary $z$ components of the wavevector $k_{z,mn}$ as well as complex angles of diffraction $\sin(\theta_{mn})$. 
These evanescent orders decrease exponentially with the distance from the grating, and hence, can be detected only at a distance less than a few wavelengths from the grating. However, these evanescent orders play a key role in some surface-enhanced grating properties and are taken into account in the theory of gratings. Evanescent orders
are essential in some special applications, such as for instance waveguide and fiber gratings. The specular order ($m = 0$) is always propagating while the others can be either propagating or evanescent. The modulations with $2\pi/K<< \lambda_0$ will produce evanescent orders for $m \neq 0$, while the modulations with $2\pi/K>> \lambda_0$ will yield a large number of propagating orders.

In the homogeneous regions, i.e., regions 1 and 3, the magnitude of the wavevectors of the backward- and forward-diffracted orders read 
$|k_{mn}|=|k_n|, \qquad \text{and}  \qquad |k''_{mn}|=|k''_n|$. The $x$ components of the diffracted wavevectors, $k_{x,mn}$ and $k''_{x,mn}$, can be deduced from the phase-matching requirements. Then, the propagating and evanescent nature of the corresponding orders will be specified based on the $k_{z,mn}$ and $k''_{x,mn}$, as follows. The real $k_{z,mn}$s and $k''_{z,mn}$s correspond to propagating orders, whereas the imaginary $k_{z,mn}$s and $k''_{z,mn}$s correspond to evanescent orders. The propagating and evanescent $m$th fields in
regions 1 and 3 are shown in Fig.~\ref{Fig:16}(c). The wavevectors in regions 1 and 3 possess magnitudes $|k_{n}|$ and $|k''_n|$, respectively. Hence, all the spatial diffraction orders for the $n$th temporal harmonic in these two regions share the same amplitude, i.e, $|k_{mn}|=|k_n|$ and $|k''_{mn}|=|k''_n|$. Semicircles with these radii are sketched in Fig.~\ref{Fig:16}(c). The allowed wavevectors in these regions must be phased matched to the boundary components of the ST diffracted order inside the grating. This is shown by the horizontal dashed lines in the figure. In the qualitative illustration in Fig.~\ref{Fig:16}(c), for the incident wave of wavevector $k_0$ and the grating with grating wavevector $K$ and temporal frequency $\Omega$, the $m = -1$ to $+2$ waves exist as propagating diffracted orders in regions 1 and 3. However, $m\leqslant-2$ and $m \geqslant +1$ will be diffracted as evanescent orders.

First, we expand the field inside the modulated medium in terms of the ST diffracted orders ($m$ and $n$) of the field in the periodic structure. This is due to the fact that the electromagnetic waves in periodic media take on the same periodicity as their host. These ST diffracted orders inside the grating are phase matched to diffracted orders (either propagating or evanescent) outside of the grating. The partial ST harmonic fields may be considered as inhomogeneous plane waves with a varying amplitude along the planar phase front. These inhomogeneous plane waves are dependent and they exchange energy back and forth between each other in the modulated grating. 

Since the electric permittivity of the grating is periodic in both space and time, with spatial frequency $K$ and temporal frequency $\Omega$, it may be expressed in terms of the double Fourier series expansion, as $n_\text{gr}^2(x,t)= \epsilon_\text{gr}(x,t)=\sum_{m}\sum_{n}  \epsilon_{mn} \exp(i (mK x  -n \Omega t))$, where $\epsilon_{mn}$ are complex coefficients of the permittivity, and $K$ and $\Omega$ are the spatial and modulation frequencies, respectively. The electric field inside the grating is expressed in terms of a sum of an infinite number of modes, i.e., $\mathbf{E}_{2}(x,z,t)=  \sum_{p} \mathbf{E}_{2,p}(x,z,t)$. Given the ST periodicity of the grating, the corresponding electric field of the $p$th mode inside the grating may be decomposed into ST Bloch-Floquet plane waves, as
\begin{subequations} 
\begin{equation}\label{eqa:Emp}
\begin{split}
\mathbf{E}_{2,p}(x,z,t)=\mathbf{\hat{y}}  \sum_{m} \sum_{n} E'_{pmn} e^{i ( k'_{x,pmn} x +k'_{z,pmn} z -\omega_{n} t)} ,
\end{split}
\end{equation} 
where 
\begin{equation}\label{eqa:Max2_c}
\begin{split}
& E'_{pmn}=\frac{(\omega_{n}/c)^2}{(k'_{x,pmn})^2 +(k'_{z,pmn})^2 }  \sum_{j}\sum_{q} \epsilon_{m-j,n-q}  E'_{pjq} 
\end{split}
\end{equation}
\begin{equation}\label{eqa:kka}
\begin{split}
k'_{x,pmn}=\left(k'_{p00}+n \frac{\Omega}{v'_\text{r}} \right) \sin\left( \tan^{-1}\left( \frac{k'_{x,p0n}}{k'_{z,p0n}}  \right) \right)+ mK 
\end{split}
\end{equation} 
\end{subequations} 
and $k'_{z,pmn}=k'_{pmn} \cos(\theta_{\text{i}})$.

Next, we determine the backward diffracted fields in region 1 and forward diffracted fields in region 3. As depicted in Fig.~\ref{Fig:16}(c), one must consider the multiple backward and forward-propagating diffracted orders that exist inside and outside of the grating. The total electric field in region 1 is the sum of the incident and the backward-traveling diffracted orders, as
\begin{equation}\label{eqa:ER}
\begin{split}
\mathbf{E}_\text{1} =\mathbf{\hat{y}} E_{0} e^{i ( k_{x,\text{i}} x +k_{z,\text{i}} z -\omega_{0} t)}+ \mathbf{\hat{y}} \sum_{m,n} E^\text{R}_{mn} e^{i ( k_{x,mn} x -k_{z,mn} z -\omega_{n} t)},
\end{split}
\end{equation} 
where $E^\text{R}_{mn}$ is the unknown amplitude of the $m$th reflected ST diffracted orders in region 1, with the wavevectors $k_{x,mn}$ and $k_{z,mn}$. The total electric field in region 3 reads $\mathbf{E}_\text{3} = \mathbf{\hat{y}} \sum_{m,n} E^\text{T}_{mn} e^{i ( k''_{x,mn} x +k''_{z,mn} z -\omega_{n} t)}$, where $E^\text{T}_{mn}$ is the amplitude of the $m$th transmitted ST diffracted order in region 3, with the wavevectors $k''_{x,mn}$ and $k''_{z,mn}$. To determine the unknown field coefficients of the backward and forward diffracted orders, $E^\text{R}_{mn}$ and $E^\text{T}_{mn}$, we enforce the continuity of the tangential electric and magnetic fields at the boundaries of the grating at $z=0$ and $z=d$.

\begin{figure}
	\begin{center}
	\includegraphics[width=0.85\columnwidth]{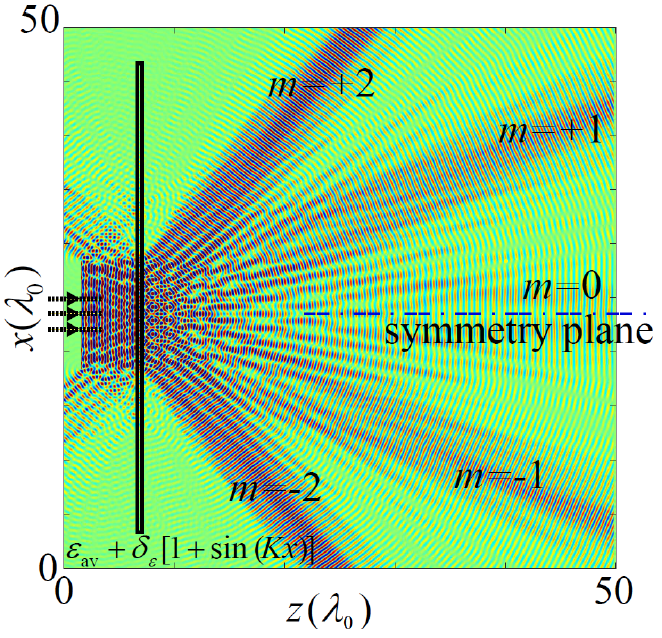}
	\caption{FDTD numerical simulation results of the $y$-component of the electric field for the diffraction from a conventional spatially periodic (static) grating ($\Omega=0$)~\cite{taravati_PRApp_2019}.}
	\label{fig:time_conv}
\end{center}
\end{figure}

\begin{figure}
	\begin{center}
\includegraphics[width=1\columnwidth]{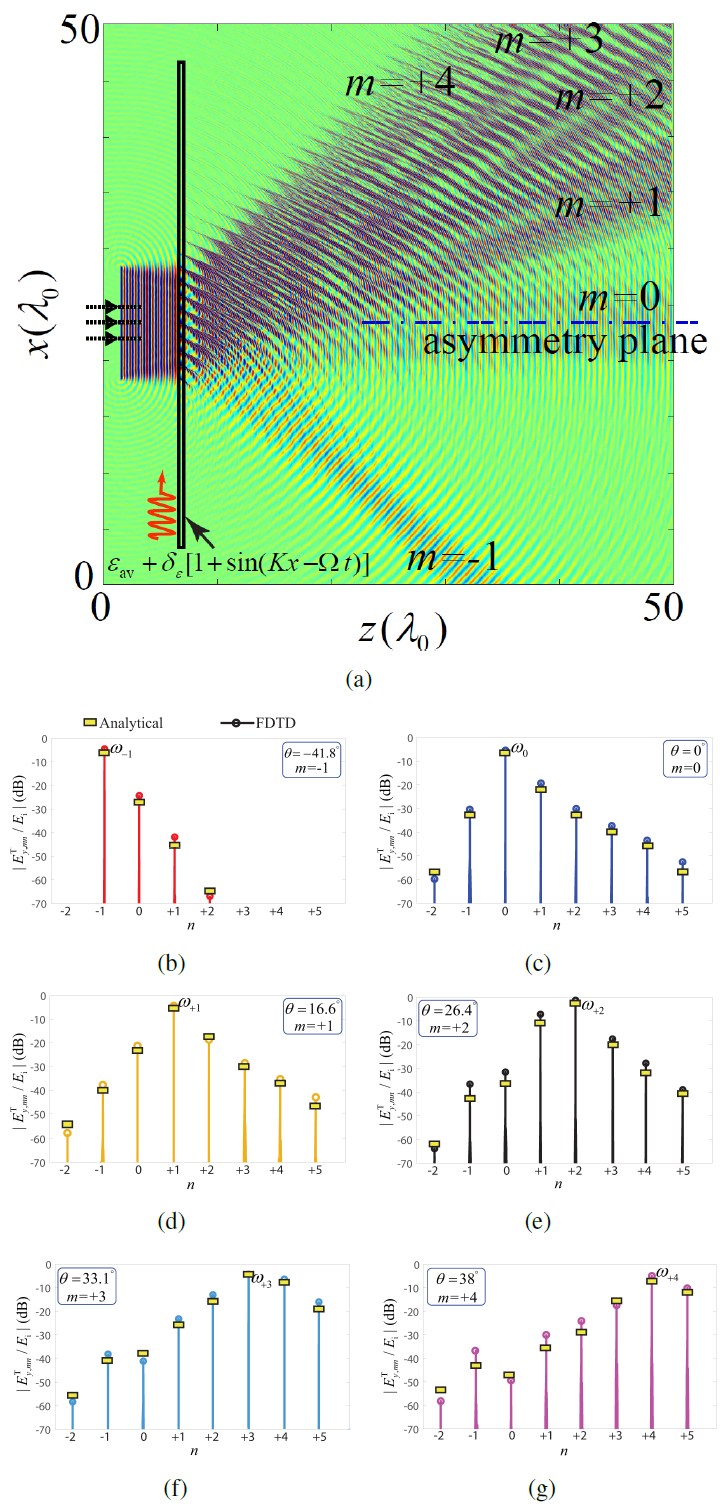}
	\caption{Analytical and FDTD simulation results for the $y$-component of the electric field, for the ST diffraction from a STP grating with $\omega_0=2\pi\times10$~GHz, where $\delta_{\epsilon}=0.5$, $\Omega=0.28\omega_0$, $K=0.4 k_0$, $d=0.8 \lambda_0$. (a) Time domain response~\cite{taravati_PRApp_2019}. (b)-(g) Frequency domain responses for $m=-1$ to $m=+4$ ST diffraction orders~\cite{taravati_PRApp_2019}.}
	\label{Fig:18}
\end{center}
\end{figure}

For the sake of comparison, we first investigate the diffraction from a conventional planar spatially periodic (static) diffraction gratings. Such a static grating assumes a sinusoidal relative electric permittivity. Figure~\ref{fig:time_conv} shows the time domain FDTD simulation results for the diffraction from a conventional spatially periodic grating with $\theta_\text{i}=0^\circ$, $\omega_0=2\pi\times10$~GHz, $\delta_{\epsilon}=0.5$, $\Omega=0$, $K=0.4 k_0$, $d=0.8 \lambda_0$. It may be seen from this figure that, for a monochromatic incident wave, all spatial diffracted orders possess the same wavelength (frequency). Another observed phenomenon is that, since the grating is "undirectional", the diffraction pattern for a normal incidence ($\theta_{\text{i}}=0$) is symmetric with respect to the $x$ axis. 
Next, we demonstrate the diffraction from a planar STP (dynamic) diffraction grating. As a particular case, which is practical and common, we study the grating with a sinusoidal relative electric permittivity in the region from $z = 0$ to $z = d$ given by $n_\text{gr}^2(x,t)= \epsilon_\text{av} +\delta_\epsilon  [1+\sin(Kx-\Omega t)]$. Figure~\ref{Fig:18}(a) shows the corresponding time domain FDTD simulation results for the diffraction from this STP grating. This figure shows that, different than the conventional case in Fig.~\ref{fig:time_conv}, the diffraction pattern of the STP grating is asymmetric with respect to the $x$ axis. Furthermore, the diffracted orders acquire different wavelengths, which correspond to different frequencies. Figures~\ref{Fig:18}(b) to~\ref{Fig:18}(g) plot the analytical and FDTD numerical simulation frequency domain responses for the $m=-1$ to $m=+4$ diffracted orders. These figures show that each diffracted spatial order includes an infinite set of temporal harmonics, $\omega_n$. 
\subsection{Effect of the grating thickness}
It is of great interest to investigate the effect of the thickness of the STP grating ($d$) on the generation of space and time diffraction orders and the grating efficiency. In general, diffraction gratings may be classified in two main categories, i.e., \textit{thin} and \textit{thick} gratings, each of which exhibiting its own angular and wavelength selectivity characteristics. The thin gratings usually result in Raman-Nath regime diffraction, where multiple diffracted orders are produced. In contrast, the thick gratings usually result in Bragg regime diffraction, where only one single diffracted order is produced. Following the procedure described in~\cite{hutley1982diffraction,gaylord1985analysis}, we characterize these two diffraction regimes, i.e., the Bragg and Raman-Nath regimes, by the dimensionless parameter 
\begin{equation}
Q_n=\frac{v_\text{r} K^2 d}{(\omega_0+n \Omega) \cos(\theta'_n) }
\end{equation}

The grating strength parameter is represented by 
\begin{equation}
\xi_n=\frac{\delta_\epsilon}{\epsilon_\text{av}}\frac{ d (\omega_0+n \Omega)}{4 v_\text{r}  \cos(\theta'_n) }
\end{equation}
for TE polarization, and
\begin{equation}
\xi_n=\frac{\delta_\epsilon}{\epsilon_\text{av}}\frac{ d (\omega_0+n \Omega) \cos(2\theta'_n)}{4 v_\text{r}  \cos(\theta'_n) }
\end{equation}
for TM polarization.
\begin{figure}
	\begin{center}
\includegraphics[width=0.9\columnwidth]{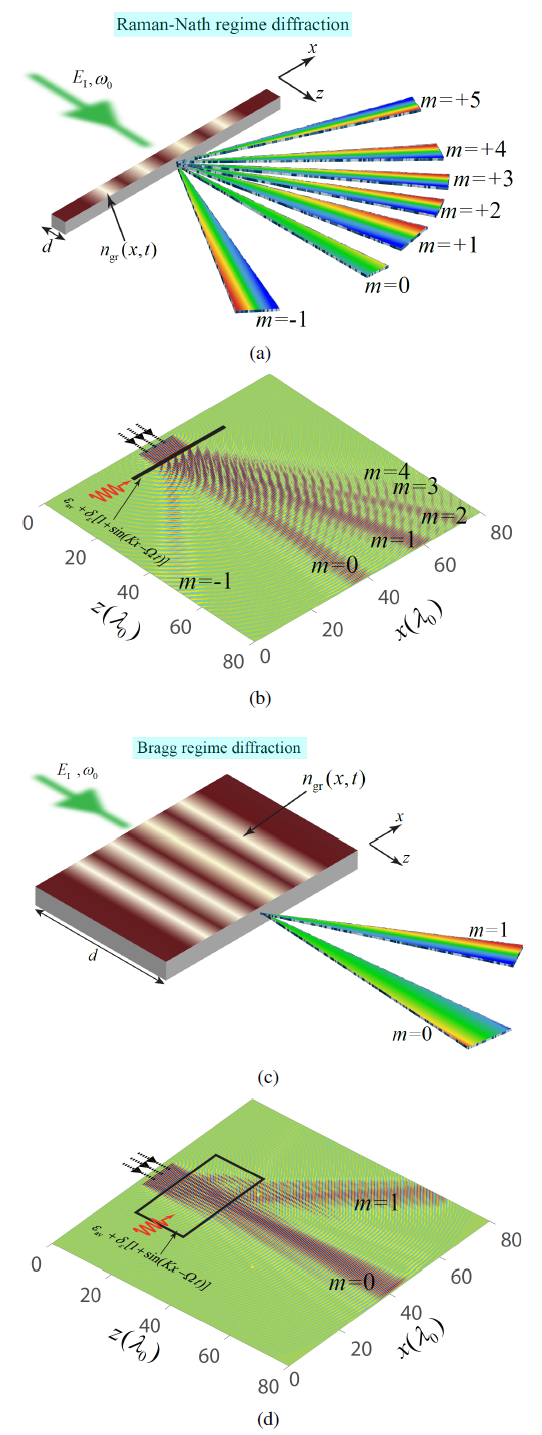}
	\caption{Two different operation regimes of STP transmissive diffraction gratings. (a) Raman-Nath regime diffraction of a thin grating, where $\Omega=0.4\omega_0$ and $K=0.4 k_0$, $\delta_{\epsilon}=0.5$ and $d=0.5 \lambda$~\cite{taravati_PRApp_2019}. (b) Bragg regime diffraction of a thick grating, where $\Omega=0.347\omega_0$ and $K=0.867 k_0$, $\delta_{\epsilon}=0.1$ and $d=16 \lambda$~\cite{taravati_PRApp_2019}.}
	\label{Fig:19}
	\end{center}
\end{figure}

\subsubsection{Raman–Nath regime}: The required condition for thin STP gratings exhibiting Raman-Nath regime
diffraction is represented by $Q_n \xi_n  \leq 1$. Thin gratings may be also characterized as gratings showing small angular and wavelength selectivity. As the incident wave is dephased (either in angle of incidence or in wavelength) from the Bragg condition, the diffraction efficiency decreases. The angular range or wavelength range for which the diffraction efficiency decreases to half of its
on-Bragg-angle value is determined by the thickness of the
grating $d$ expressed as a number of grating periods $\Lambda=2\pi/K$. For a
thin grating this number is reasonably chosen to be $Kd   \leq 20\pi$. Figure~\ref{Fig:19}(a) shows a generic representation of the Raman-Nath regime diffraction in STP transmissive diffraction gratings, for normal incidence ($\theta_\text{i}=0^\circ$), $\omega_0=2\pi\times10$~GHz, $\Omega=0.4\omega_0$ and $K=0.4 k_0$, $\delta_{\epsilon}=0.5$ and $d=0.5 \lambda$. Figure~\ref{Fig:19}(b) shows the numerical simulation results for Raman-Nath regime diffraction of the STP grating in Fig.~\ref{Fig:19}(a). Following the procedure described in~\cite{hutley1982diffraction,gaylord1985analysis} (for conventional spatially periodic gratings), for a thin transmissive STP grating operating in the Raman-Nath regime, the diffraction efficiency reads $\eta_{mn}=J_m^2(2\xi_n)$, where $P_{mn}$ and $P_\text{inc}$ are the diffracted and incident powers, respectively, and where $J$ represents the, integer-order, ordinary Bessel function of the first kind.

\subsubsection{Bragg regime}: The Bragg regime diffraction may be achieved in thick gratings. Thick gratings are capable of exhibiting strong angular and wavelength selectivity. A relatively small change in the angle of incidence from the Bragg angle or a relatively small change in the wavelength at the Bragg angle may result in a relatively strong dephasing, which in turn, decreases the diffraction efficiency. Thick grating behavior occurs when $Kd  \geq 20\pi$. Figure~\ref{Fig:19}(c) shows a generic representation of the Bragg regime diffraction in STP transmissive diffraction gratings, for normal incidence ($\theta_\text{i}=0^\circ$), $\omega_0=2\pi\times10$~GHz, $\Omega=0.4\omega_0$ and $K=0.4 k_0$, $\delta_{\epsilon}=0.5$ and $d=0.5 \lambda$. Figure~\ref{Fig:19}(d) shows the numerical simulation results for Bragg regime diffraction of the STP grating in Fig.~\ref{Fig:19}(c). Following the procedure described in~\cite{hutley1982diffraction,gaylord1985analysis} (for conventional spatially periodic gratings), for a thick transmissive STP grating operating in the Bragg regime, the diffraction efficiency reads $\eta_{1n}=\sin^2(2\xi_n)$.

\subsection{Asymmetric and Nonreciprocal Diffractions}\label{sec:nonr_asymm}
\subsubsection{Transmissive STP Grating}
\begin{figure*}
		\begin{center}
\includegraphics[width=2\columnwidth]{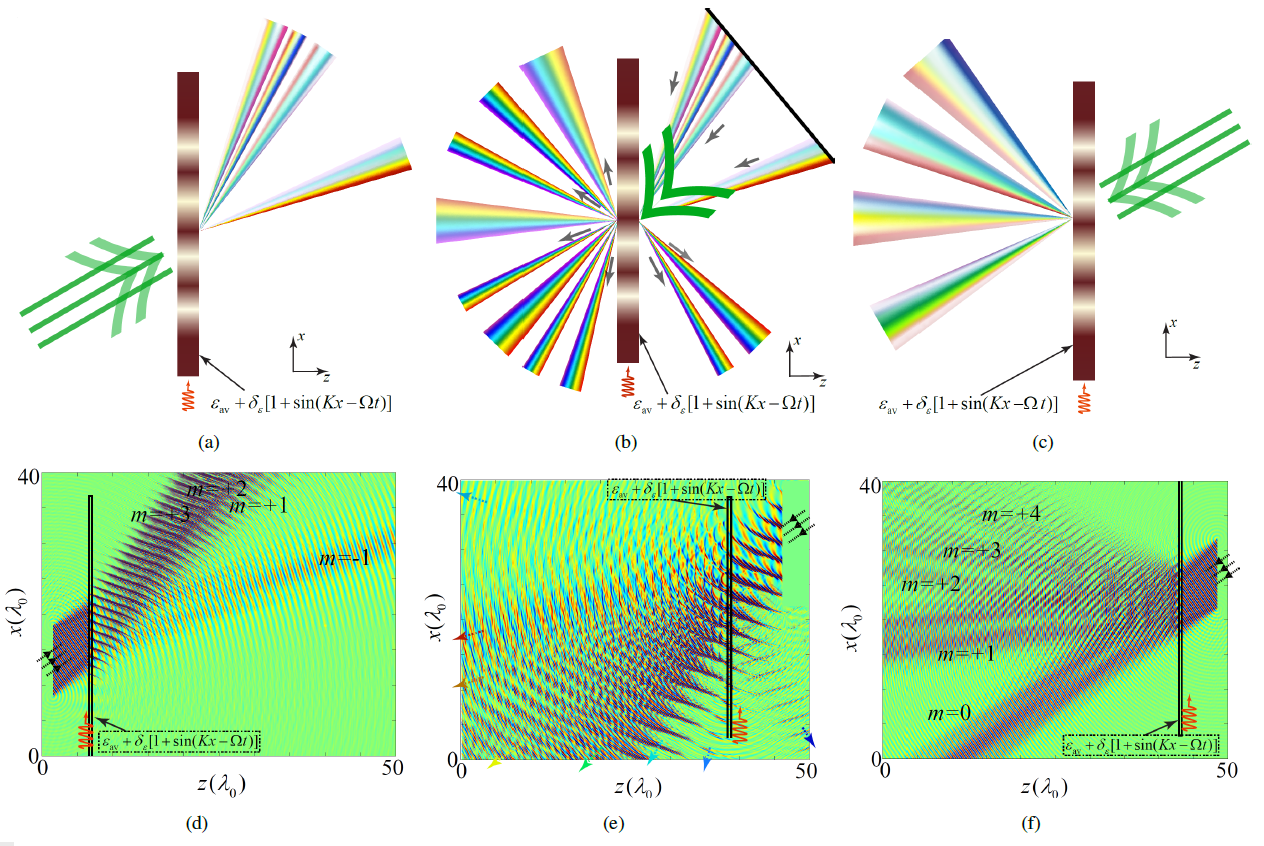}
	\caption{Nonreciprocal and asymmetric wave diffraction from a transmissive STP grating with sinusoidal STM permittivity, i.e., $\epsilon(x,t)=\epsilon_\text{av}+\delta_\epsilon \sin(K x-\Omega t)$, where $\theta_\text{i}=35^\circ$, $\omega_0=2\pi\times10$~GHz, $\delta_{\epsilon}=0.5$, $\Omega=2\pi\times4$~GHz and $d=0.8 \lambda_0$. (a) and (d) Forward problem~\cite{taravati_PRApp_2019}. (b) and (e) Backward problem for demonstration of nonreciprocal wave diffraction~\cite{taravati_PRApp_2019}. (c) and (f) Backward problem for demonstration of asymmetric wave diffraction~\cite{taravati_PRApp_2019}.}  
	\label{Fig:20}
	\end{center}
\end{figure*}

Figure~\ref{Fig:20}(a) illustrates a particular example, where a $+z$-propagating incident field (forward problem) obliquely impinges on a transmissive STP grating. Figure~\ref{Fig:20}(d) shows the FDTD numerical simulation result of the transmissive diffraction by the STP diffraction grating in Fig.~\ref{Fig:20}(a) with $\theta_\text{i}=35^\circ$, $\omega_0=2\pi\times10$~GHz, where $\delta_{\epsilon}=0.5$, $\Omega=2\pi\times4$~GHz, $d=0.8 \lambda_0$. As expected, the diffracted ST orders possess different wavelengths and different amplitudes. Figures~\ref{Fig:20}(b) to~\ref{Fig:20}(f) present the simulation results for the nonreciprocity and angle-asymmetric operation of the transmissive STP diffraction gratings. It may be seen from Figs.~\ref{Fig:20}(b) and~\ref{Fig:20}(e) that the output of the backward problem is completely different than the incident wave of the forward problem, showing a nonreciprocal wave diffraction. This asymmetric diffraction test is depicted in Fig.~\ref{Fig:20}(c), and the corresponding time domain response is given in Fig.~\ref{Fig:20}(f). Comparing the numerical simulation results in Figs.~\ref{Fig:20}(d) and~\ref{Fig:20}(f), we see that the STP grating introduces completely different diffraction patterns for forward and backward incidence, which includes difference in the angle of diffraction and amplitude of the diffracted fields.
\subsubsection{Reflective STP Grating}
\begin{figure*}
	\begin{center}
\includegraphics[width=2\columnwidth]{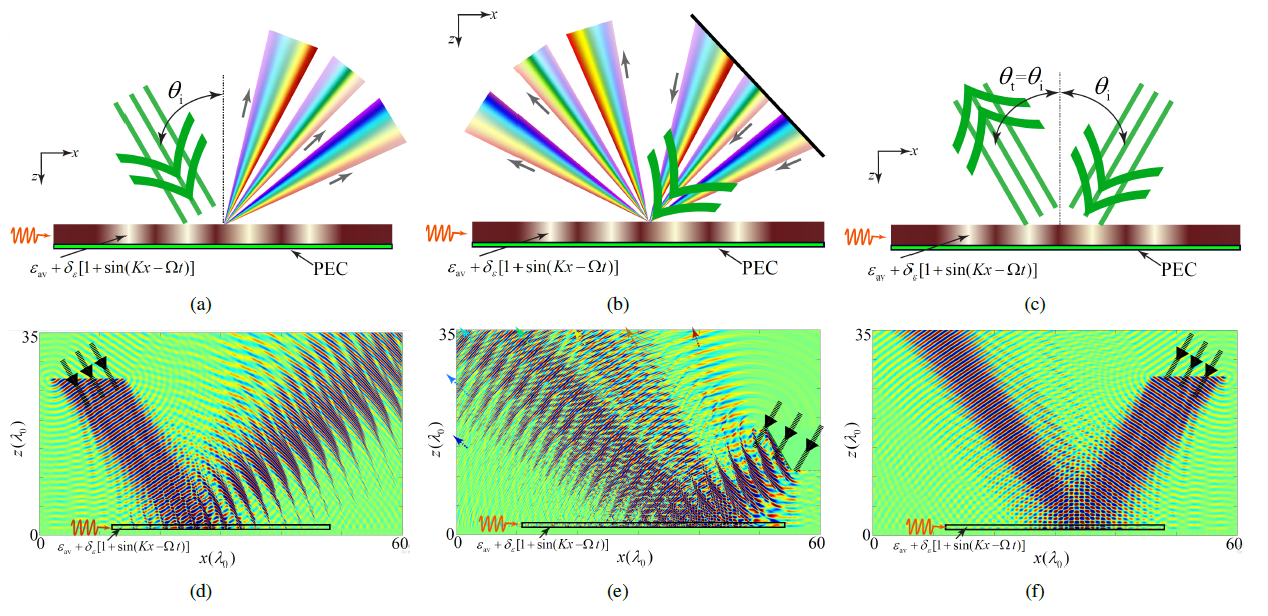}
	\caption{Nonreciprocal and angle-asymmetric ST diffraction of a reflective STP diffraction grating with a $+x$-traveling STM electric permittivity, i.e., $\epsilon(x,t)=\epsilon_\text{av}+\delta_\epsilon [1+\sin(K x-\Omega t)]$, where $\delta_{\epsilon}=0.5$, $\Omega=0.4 \omega_0$, $d=0.8 \lambda_0$. (a) and (d) Forward wave incidence~\cite{taravati_PRApp_2019}. (b) and (e) Backward wave incidence for nonreciprocal diffraction demonstration~\cite{taravati_PRApp_2019}. (c) and (f) Backward wave incidence for angle-asymmetric demonstration~\cite{taravati_PRApp_2019}.}  
	\label{Fig:21}
\end{center}
\end{figure*}

Figures~\ref{Fig:21}(a) to~\ref{Fig:21}(f) show the nonreciprocal and angle-asymmetric responses of reflective STP diffraction gratings. It may be seen from Fig.~\ref{Fig:21}(b) and~\ref{Fig:21}(e) that the output of the backward problem is totally different than (the spatial inversion of) the incident wave of the forward problem, which demonstrates strong nonreciprocity of the reflective STP grating. Comparing the results of the forward and backward incidence, shown in Figs.~\ref{Fig:21}(d) and~\ref{Fig:21}(f), respectively, one may obviously see that the reflective diffraction by the grating is completely angle-asymmetric. Such an asymmetric reflective diffraction includes asymmetric angles of diffraction and unequal amplitudes of the diffracted orders.

\section{Diffraction code multiple access system}\label{sec:app}
The proposed STP grating offers unique properties that can be utilized for the realization of new types of electromagnetic devices and operations, such as for instance, nonreciprocal beam shaping and beam coding, multi-functionality antennas, tunable and nonreciprocal beam steering, enhanced resolution holography, multiple images holography, illusion cloaking, etc.

\begin{figure}
	\begin{center}
\includegraphics[width=1\columnwidth]{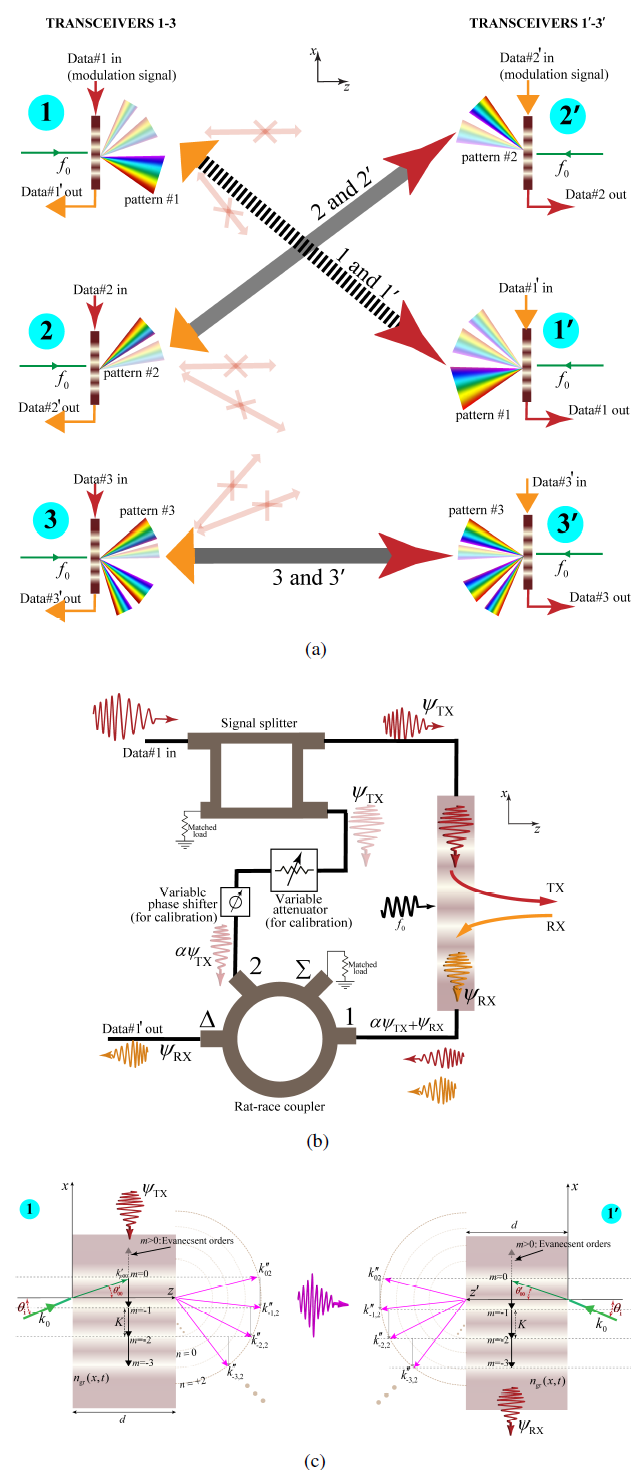}
	\caption{ST diffraction code multiple access (STDCMA) system. (a) Schematic representation~\cite{taravati_PRApp_2019}. (b) Full-duplex operation mechanism~\cite{taravati_PRApp_2019}. (c) Wavevector diagram of a particular transceiver pair~\cite{taravati_PRApp_2019}.}
	\label{Fig:22}
	\end{center}
\end{figure}

Figure~\ref{Fig:22}(a) presents an original application of the STP diffraction grating to wireless communications. Such a communication system is hereby called ST diffraction code multiple access (STDCMA) system. In the example provided in Fig.~\ref{Fig:22}(a), we consider three pairs of transceivers (in practice one may consider more pairs of transceivers). In such a scenario, only the transceiver pairs that share the same ST diffraction pattern can communicate. Each diffraction pattern is attributed to the properties of the grating ST modulation, i.e., the input frequency, where the input data (message) plays the role of the modulation signal. For a specified input data (modulation signal), a unique diffraction pattern is created. In the particular example in Fig.~\ref{Fig:22}(a), the transceiver pairs that are allowed to communicate are $1$ and $1'$, $2$ and $2'$, and $3$ and $3'$, so that the transceivers $2'$ and $3'$ ($2$ and $3$) are incapable of retrieving the data sent by the transceiver $1$ ($1'$), and the transceivers $1'$ and $3'$ ($1$ and $3$) are incapable of retrieving the data sent by the transceiver $2$ ($2'$), and so forth. Each communication pair shares a certain ST diffraction \textit{pattern}. Each diffraction pattern can be created by certain ST modulation parameters, e.g. $\delta_{\epsilon}$, $\epsilon_\text{av}$, the $K/\Omega$ ratio, and the grating thickness $d$. Since the radiation pattern provided by a STP diffraction grating is very diverse and is very sensitive to the ST modulation parameters, an optimal isolation between the transceivers can be achieved by proper design of the diffraction patterns.

Such a multiple access scheme is endowed with full-duplex operation, thanks to the unique nonreciprocity provided by the properties of a STP diffraction grating. Figure~\ref{Fig:22}(b) depicts the architecture of a STP-diffraction-grating-based transceiver in the STDCMA system in Fig.~\ref{Fig:22}(a). Such an architecture is composed of a STP diffraction grating illuminated by an incident wave with frequency $f_0$. In the transmit mode (TX), the grating is modulated by the input data denoted by $\psi_\text{TX}$ which is injected to the grating from the top and travels in the $-x$ direction. In the receive mode (RX), the incoming wave (which includes a set of ST diffraction orders) impinges on the grating and while interacting with the incident wave with frequency $f_0$, yields a $-x$ traveling wave inside the grating, denoted by $\psi_\text{RX}$. We shall stress that traveling of the $\psi_\text{RX}$ signal in the $-x$ direction is enforced by proper design of the grating, which will be explained later.

As we see in Fig.~\ref{Fig:22}(b), the signal wave at the receiver port is composed of the received signal ($\psi_\text{RX}$) plus a portion of the input data of the transmission mode ($\alpha \psi_\text{TX}$). To ensure complete cancellation of the $\psi_\text{TX}$ in the receiver port, we may use the circuit in the left side of Fig.~\ref{Fig:22}(b). This circuit is composed of a signal splitter that provides a sample from the input data of the transmit mode ($\psi_\text{TX}$), a variable attenuator and a variable phase shifter for calibration purposes to provide $\alpha \psi_\text{TX}$. Then, the signal wave at the receiver port, i.e., $\psi_\text{RX}+\alpha \psi_\text{TX}$, will be subtracted from the calibrated sample signal, that is $\alpha \psi_\text{TX}$, by a rat-race coupler. Thus, the signal at the difference port of the rat-race coupler is the desired received signal $\psi_\text{RX}$. It is worth mentioning that the calibration of the architecture can be performed by disconnecting the RX port from port-1 of the rat-race coupler, connecting port-1 to a match load, and then seeking for a null at the difference port of the rat-race coupler by adjusting the variable attenuator and variable phase shifter, so that $\psi_\text{TX}$ is completely canceled out at the difference port of the rat-race coupler.

An elegant feature of the transceiver scheme in Fig.~\ref{Fig:22}(b) is that the diffraction grating is used as the receiver (as well as the transmitter), where the received signal wave acts as the modulation signal (specifies the $K$ and $\Omega$ parameters) instead of the incident field. The key reason for the duplexing operation is that, inside the grating the wave can only flow downstream. Figure~\ref{Fig:22}(c) shows how the full-duplex operation is achieved by proper design of the diffraction grating, where only negative diffraction orders, i.e., $-x$ propagating orders, are generated, while positive diffraction orders, that is $+x$ propagating orders, are evanescent. This way, we ensure that inside the grating, all the diffraction orders are traveling in the $-x$ direction, in both the transmit and receive modes. In Fig.~\ref{Fig:22}(c), transceiver $1$ operates in the transmit mode, where the input data ($\psi_\text{TX}$) is injected to the grating from the top and while interacting with the incident wave with the wavenumber $k_0$, generates a number of nonpositive diffraction orders, i.e., $m \geq 0$. In the right side of Fig.~\ref{Fig:22}(c), transceiver $1'$ receives the diffracted orders by transceiver $1$, so that the resultant wave inside $\psi_\text{RX}$ exits the grating from the bottom port of the grating, as all the diffraction orders can only travel in the $-x$ direction.

\section{Unidirectional Beam Splitter}
Beam splitters are quintessential elements of communication systems. In spite of the immense scientific attempts for the realization of efficient beam splitters, beam splitters are restricted to reciprocal response and suffer from substantial transmission loss. As a consequence, the resource requirements of the overall system, including demand for high power microwave sources and isolators, will be increased. Here, we introduce a one-way, tunable and highly efficient beam splitter and amplifier based on coherent electromagnetic transitions through the oblique illumination of STM structures. The contributions in this regard are as follows. In contrast to conventional beam splitters which are restricted to reciprocal response with more than 3 dB insertion loss, the proposed STM beam splitter is capable of providing nonreciprocal response with transmission gain. It can be also used in antenna applications, where the transmitted and received waves are engineered appropriately. Moreover, the angle of transmission and the amplitude of the transmitted beams depend on the ST modulation parameters. Hence, the ST modulation parameters provide the leverage for achieving the desired angle of transmission for the two output beams of the STM beam splitter. In addition, unequal power division between the output beams can be achieved by varying the ST modulation parameters.

\begin{figure*}
	\begin{center}
\includegraphics[width=2\columnwidth]{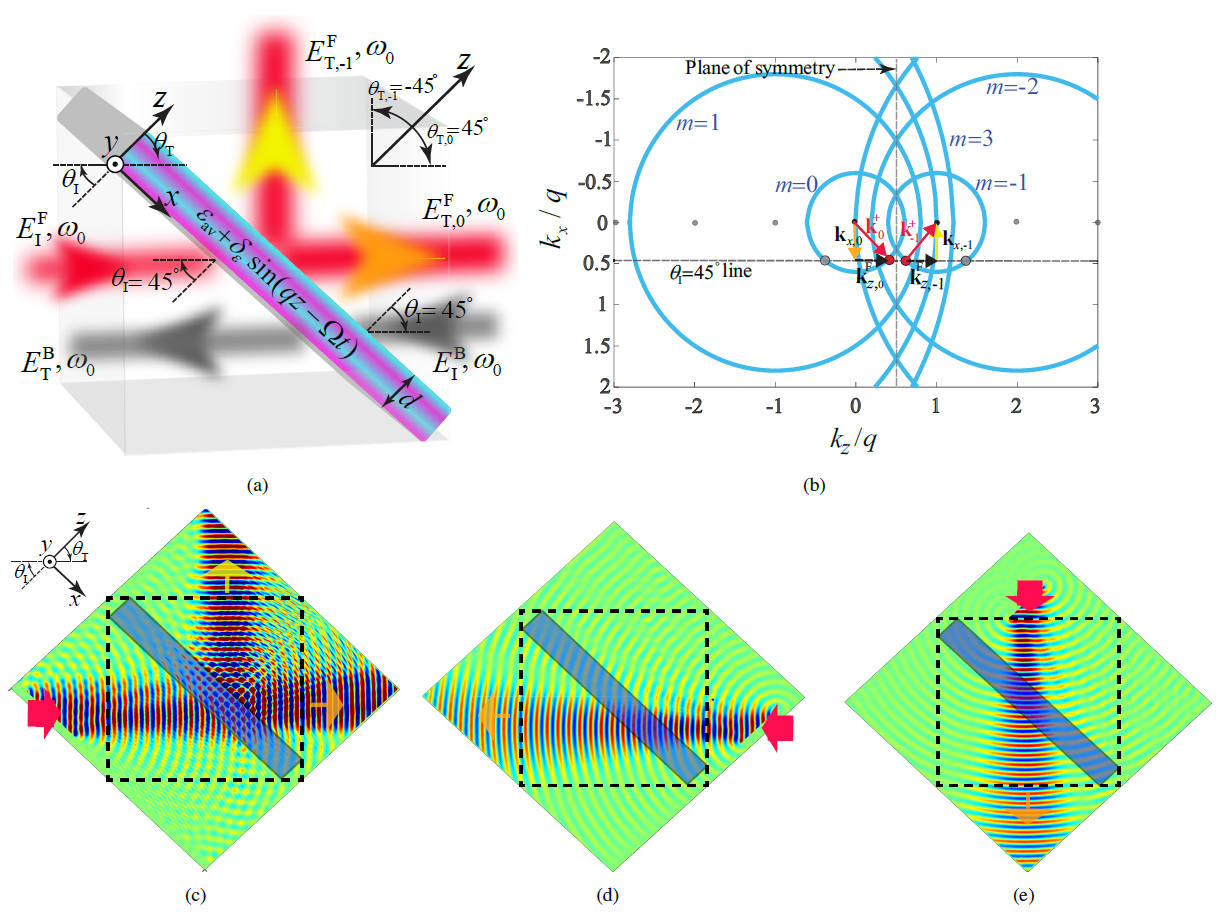}
		\caption{One-way beam splitting by a STM metasurface. (a) Schematic~\cite{Taravati_Kishk_PRB_2018}.  (b)~Isofrequency diagram at $\omega=\omega_0$ composed of an infinite set of circles centered at $(k_z/q,k_{x,\text{i}}/q)=(-n,0)$ with radius $\varGamma(0.5+n)$~\cite{Taravati_Kishk_PRB_2018}.
		(c)-(e) FDTD numerical simulation for the wave incidence to the metasurface for the forward wave incidence to the metasurface, from the left ($\theta_\text{I}=45^\circ$),~from the right ($\theta_\text{I}=45^\circ$), and from the top ($\theta_\text{I}=-45^\circ$), respectively~\cite{Taravati_Kishk_PRB_2018}.}
		\label{Fig:23}
	\end{center}
\end{figure*}

Figure~\ref{Fig:23}(a) sketches the nonreciprocal beam transmission and splitting in a STM metasurface. By appropriate design of the band structure, that is, the ST modulation format and its associated temporal and spatial modulation frequencies, unidirectional energy and momentum exchange between the incident wave-under angle of incidence and transmission $\theta_{\text{i}}=\theta_{\text{T},0}=45^\circ$ and temporal frequency $\omega_0$- to the first lower STH-under angle of transmission $\theta_{\text{T},-1}=-45^\circ$ and temporal frequency $\omega_0$- will occur. We assume the incident electric field in Eq.~\eqref{eqa:Ei1} under the angle of incidence $\theta_{\text{i}}=45^\circ$ impinges to the periodic STM metasurface. 

The STM metasurface assumes a sinusoidal ST-varying permittivity with the temporal modulation frequency of $\Omega=2 \omega_0$, and the spatial modulation frequency of $q= 2 k_0/\varGamma$. Here, $k_{z,n}=\beta_{0}+nq$ and the temporal frequency $\omega_n=(1+2n) \omega_0$, with $\beta_{0}$ being the unknown spatial frequency of the fundamental harmonic. The unknowns of the electric field, that is, $A_n$ and $\beta_0$, will be found through satisfying Maxwell's equations.

The transmission angle of the $m$th transmitted STH, $\theta_{\text{T},n}$, satisfies the Helmholtz relation as $k_{0}^2 \sin^2(\theta_{\text{i}})+k_{n}^2 \cos^2(\theta_{\text{T},n})=k_{n}^2$, where $k_{n}=\omega_n/v_\text{b}$ denotes the wavenumber of the $n$th transmitted STH outside the STM metasurface. The angle of transmission $\theta_{\text{T},n}$ reads
\begin{equation}\label{eqa:mm}
	\theta_{\text{T},n}=\sin^{-1} \left(\frac{k_{x,\text{i}}}{k_n} \right)=\sin^{-1} \left(\frac{\sin(\theta_\text{i})}{1+2n} \right).
	\end{equation} 
which demonstrates the spectral decomposition of the transmitted wave. Consequently, the fundamental STH, $n=0$, and the first lower STH, $n=-1$, with equal temporal frequency $\omega_0$, will be respectively transmitted under the angles of transmission of $\theta_{\text{T},0}=\theta_{\text{i}}=45^\circ$ and $\theta_{\text{T},-1}=-\theta_{\text{i}}=-45^\circ$ so that they are transmitted under $90^\circ$ angle difference, presenting the desired beam splitting. 

Figure~\ref{Fig:23}(b) presents the isofrequency dispersion diagram. This diagram is formed by $2N+1$ periodic set of double cones (here, only $m=0$ and $n=-1$ harmonics are shown), each of which representing a STH, with apexes at $k_{x,\text{i}} = 0$, $k_z = -n q$ and $\omega=-2n \omega_0$, and the slope of $v_\text{m}$ with respect to $k_z-k_{x,\text{i}}$ plane. Consider oblique incidence of a wave, representing the fundamental harmonic $n=0$ with temporal frequency $\omega_0$, propagating along the [$+x$,$+z$] direction. It is characterized by $x$- and $z$-components of the spatial frequency, $\textbf{k}_{x}=\mathbf{\hat{x}} k_{x,\text{i}}$ and $\textbf{k}_{z}^\text{F}=\mathbf{\hat{z}} k_z$. The wave impinges to the medium under the angle of incidence $\theta_\text{I}=45^\circ$ and excites an infinite number of (we truncate it to $2N+1$) STH waves, with different spatial and temporal frequencies of $[k_{x,\text{i}},k_{z,n}]$ and $\omega_{n}$. However, interestingly, the first lower STH $n=-1$ offers similar characteristics as the fundamental harmonic, that is, the identical temporal frequency of $\omega_{0}$ and identical $z$-component of the spatial frequency of $\textbf{k}_{z,-1}^\text{F}=\textbf{k}_{z,0}^\text{F}$, but opposite $x$-component of the spatial frequency of  $\textbf{k}_{x,-1}=-\textbf{k}_{x,0}$. Hence, $m=-1$ harmonic propagates along [$-x$,$+z$] direction. In general the $x$-component of the $n$th STH reads  $\textbf{k}_{x,n}=-\textbf{k}_{x,-n-1}$). Moreover, since $\omega_{n}=\omega_{-n-1}$, the undesired STHs acquire temporal frequency of $2n\omega_0$, and are far away from the fundamental harmonic. Thus, most of the incident energy is residing in $n=0$ and $n=-1$ harmonics, both at $\omega_{0}$, respectively transmitted under $\theta_\text{T,0}=\theta_\text{I}$ and $\theta_\text{T,-1}=-\theta_\text{I}$ transmission angles with $2\theta_\text{I}$ angle difference.

Figure~\ref{Fig:23}(c) shows the numerical simulation results for the forward oblique wave incidence to the metasurface, shown in Fig.~\ref{Fig:23}(a), with $\epsilon_\text{r}=1$, $\delta_{\epsilon}=0.2$, $\varGamma=1.2$, $d=3\lambda_0=3\times2\pi/k_0$, $\theta_\text{I}=45^\circ$ and $\omega_\text{0}=3$~GHz. It may be seen from this figure that an efficient beam splitting with significant transmission gain is achieved in the forward direction. Figures~\ref{Fig:23}(d) and~\ref{Fig:23}(e) provide the results for the oblique wave incidence from the right side and top, respectively, corresponding to $\theta_\text{I}=45^\circ$ and $\theta_\text{I}=-45^\circ$. The presented analytical and numerical results demonstrate that the dynamic beam splitter provides a perfect nonreciprocal beam splitting, in the lack of beam tilting. In contrast to conventional passive beam splitters, here the beam splitting is achieved for a non-collimated beam. It may be shown that by changing the modulation parameters, i.e., $\varGamma$, $\theta_\text{I}$ and $\epsilon_\text{av}$, tunable transmission angles, unequal splitting ratio and unequal angles of transmission will be achieved. 
Given the weak transition of energy and momentum from the fundamental STH $n=0$ to higher order STHS except $n=-1$, the electric field inside the STM metasurface can be represented based on the superposition of the aforementioned two STHs, i.e., $n=0$ and $n=-1$. The electric field is then defined by
\begin{align}\label{eqa:1}
\begin{split}
E_\text{S}(x,z,t)=&a_{0}(z) e^{-i \left(k_{x,\text{i}} x+k_z z -\omega_0 t \right)}\\&+a_{-1}(z) e^{i \left(-k_{x,\text{i}} x+(q -k_z)z -\omega_0 t \right)},
\end{split}
\end{align}
where $a_{0}(z)$ and $a_{-1}(z)$ are the unknown field coefficients. We shall stress that, here the field coefficients are $z$-dependent since they include both the amplitude and the change in the spatial frequency (wavenumber) introduced by the ST modulation. The coupled differential equation for the field coefficients reads 
\begin{subequations}
	\begin{gather}\label{eqa:coupl}
	\frac{d }{d z}  \begin{bmatrix} a_{0}(z) \\ a_{-1}(z) \end{bmatrix}
	=
	\begin{bmatrix}
	M_0 & C_0 \\ C_{-1}& M_{-1}
	\end{bmatrix}
	\begin{bmatrix} a_{0}(z) \\ a_{-1}(z) \end{bmatrix},
	\end{gather}
	where
	\begin{align}
	M_0&=\frac{i k_0^2}{2k_z} (\epsilon_\text{av}-\epsilon_\text{r} ),\nonumber \\
	M_{-1}&=\frac{ik_0^2}{2(k_z-q)} 
	\left[\epsilon_\text{av}-\epsilon_\text{r}\frac{k_{x,\text{i}}^2+(q-k_z)^2}{k_0^2}  \right],\nonumber \\
	C_0&=i\frac{\delta k_0^2}{4k_z}, \quad \text{and} \quad
	C_{-1}=i\frac{\delta k_0^2}{4(k_z-q)}.
	\end{align}
\end{subequations}
\indent The solution to the coupled differential equation in~\eqref{eqa:coupl} is given by~\cite{Taravati_Kishk_PRB_2018}
\begin{subequations}\label{eqa:coup_sol}
	\begin{align}
	a_{0}(z)=\frac{E_0}{2 \varDelta} &\bigg( (M_0-M_{-1}+\varDelta ) e^{\frac{M_0+M_{-1}+\varDelta}{2}z}  \\ \nonumber
	&- (M_0-M_{-1}-\varDelta) e^{\frac{M_0+M_{-1}-\varDelta}{2}z}  \bigg),
	\end{align}
	\begin{align}
	a_{-1}(z)=\frac{E_0 C_{-1}}{ \varDelta} \left( e^{\frac{M_0+M_{-1}+\varDelta}{2}z}  - e^{\frac{M_0+M_{-1}-\varDelta}{2}z}  \right),
	\end{align}
\end{subequations}
where $\varDelta=\sqrt{(M_0-M_{-1})^{2}+4 C_0C_{-1}}$. For a given ST modulation ratio $\varGamma$, the field coefficients in Eq.~\eqref{eqa:coup_sol} acquire different forms. In general, ST modulation is classified into three categories, i.e., subluminal ($0<\varGamma<1$ or $v_\text{m}<v_\text{b}$), luminal ($\varGamma\rightarrow 1$ or $v_\text{m}\rightarrow v_\text{b}$), and superluminal ($\varGamma>1$ or $v_\text{m}>v_\text{b}$). 
\subsection{Subluminal and Superluminal ST Modulations}
\begin{figure}
	\begin{center}
\includegraphics[width=1\columnwidth]{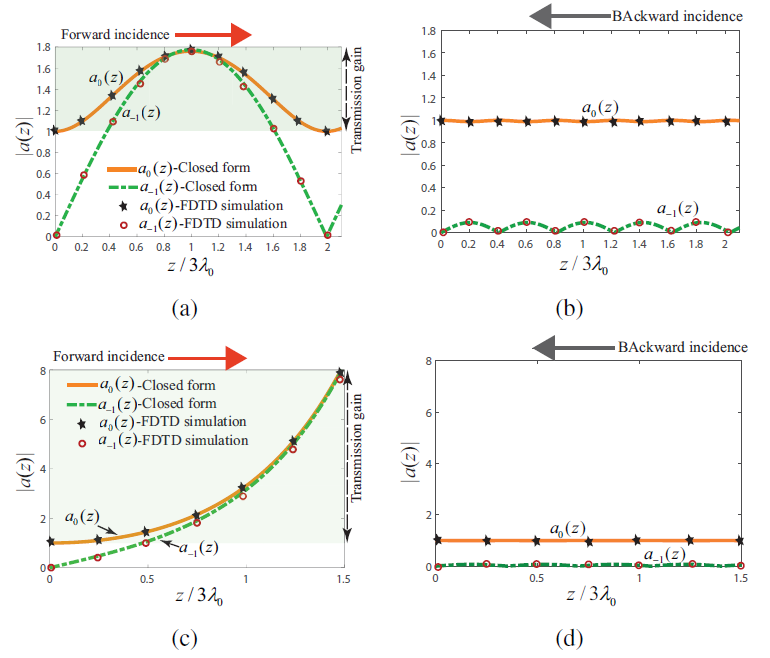}
		\caption{Closed-form solution results and the FDTD numerical simulation results for the $z$-dependent absolute field coefficients in Eq.~\eqref{eqa:1}, i.e., $a_{0}(z)$ and $a_{-1}(z)$, inside the  and luminal STM beam splitters. (a) and (b) Forward and backward wave incidence for the superluminal regime ($\varGamma=1.2$ and $\delta_\epsilon=0.28$), respectively~\cite{Taravati_Kishk_PRB_2018}. (c) and (d) Forward and backward wave incidence for the luminal regime ($\varGamma=1$ and $\delta_\epsilon=0.28$), respectively~\cite{Taravati_Kishk_PRB_2018}.}
		\label{Fig:24}
	\end{center}
\end{figure}

\indent Considering $\epsilon_\text{av}=\epsilon_\text{r}$, the $a_{0}(z)$ and $a_{-1}(z)$ in Eq.~\eqref{eqa:coup_sol} would be a periodic sinusoidal function with respect to $z$, if $\varDelta=\sqrt{(M_0-M_{-1})^{2}+4 C_0C_{-1}}$ is imaginary, i.e., $(M_0-M_{-1})^{2}+4 C_0C_{-1}<0$. By solving this, we achieve an interval for the luminal ST modulation, that is, 
\begin{equation}\label{eqa:sonic}
\varGamma_\text{sub} <\frac{1 }{\sqrt{\epsilon_\text{av} +\delta_\epsilon }}
\leq\varGamma_\text{lum}\leq
\frac{1 }{\sqrt{\epsilon_\text{av} -\delta_\epsilon }}<\varGamma_\text{sup},
\end{equation}
where $\varGamma_\text{sub}$, $\varGamma_\text{lum}$ and $\varGamma_\text{sup}$ are ST velocity ratio for subluminal, luminal and superluminal ST modulations, respectively. The interval for luminal ST modulation is called sonic regime in analogy with the sonic boom effect in acoustics, where an airplane travels with the same speed or faster than the speed of sound. It should be noted that the luminal ST modulation interval in Eq.~\eqref{eqa:sonic} is exactly the same as the one achieved from the exact analytical solution~\cite{Oliner_PIEEE_1963,Taravati_PRB_2017,Taravati_PRAp_2018}.\\
\indent Figure~\ref{Fig:24}(a) plots the closed form and FDTD numerical simulation results for the absolute electric field coefficient inside the metasurface, with the wave incidence from the left side (forward incidence), considering superluminal ST modulation of $\varGamma=1.2$ and $\delta_\epsilon=0.28$. It is seen from this figure that both $a_{0}(z)$ and $a_{-1}(z)$ possess periodic sinusoidal form and exhibit a substantial transmission gain at $z=3 \lambda_0$. Such a transmission gain is tuned through the variation of $\varGamma$ and $\delta_\epsilon$. This result is consistent with the transmission gain achieved in the FDTD numerical simulation results in Figs.~\ref{Fig:23}(c). The coherence length $l_\text{c}$, where both $a_{0}(z)$ and $a_{-1}(z)$ acquire their maximum amplitude is found as~\cite{Taravati_Kishk_PRB_2018}
\begin{align}\label{eq:iii}
l_\text{c}=\pi \left( \left[ \frac{ k_0^2[\epsilon_\text{av}-\epsilon_\text{r}]/k_z-q }{(\varGamma-2)/(\varGamma-1)}    \right]^2 +\frac{\delta^2 k_0^4}{4k_z (k_z-q)}  \right)^{-1}.
\end{align}\\
\indent Figure~\ref{Fig:24}(b) plots the result for the superluminal STM metasurface in Fig.~\ref{Fig:24}(a), except for wave incidence from the right side (backward incidence). It may be seen from this figure that, in contrast to the forward wave incidence where a substantial exchange of the energy and momentum between the $m=0$ and $m=-1$ STHs are achieved, here the incident wave passes through the metasurface with negligible alteration and minor transition of energy and momentum to the $m=-1$ STH. This is obviously in agreement with the nonreciprocal response presented in Figs.~\ref{Fig:23}(c),~\ref{Fig:23}(d) and~\ref{Fig:23}(e).
It may be shown that the for the luminal ST modulation, where $\varGamma\rightarrow1$, the field coefficients in Eq.~\eqref{eqa:coup_sol}, $a_{0}(z)$ and $a_{-1}(z)$, acquire pure real (or complex) forms. This yields exponential growth of the electric field amplitude along the STM metasurface. Hence, considering $\varGamma=1$, the total electric field inside the STM metasurface reads
\begin{align}\label{eqa}
E_\text{S}(x,z,t)&|_{\varGamma=1}=E_0 \cosh\left(\frac{\delta k_0^2}{4k_z}z\right) e^{-i \left(k_{x,\text{i}} x+k_z z -\omega_0 t \right)}\\ \nonumber &-i \frac{\delta k_0^2}{2 k_z}E_0 \sinh\left(\frac{\delta k_0^2}{4k_z}z\right) e^{i \left(-k_{x,\text{i}} x+(q -k_z)z -\omega_0 t \right)}.
\end{align}
\indent Figure~\ref{Fig:24}(c) plots the closed form and FDTD numerical simulation results for the absolute value of the electric field coefficients $a_{0}(z)$ and $a_{-1}(z)$ inside the luminal ($\varGamma=1$ and $\delta_\epsilon=0.28$) STM metasurface for forward wave incidence. It may be seen from this figure that both $a_{0}(z)$ and $a_{-1}(z)$ possess a non-periodic exponentially growing profile and exhibit a substantial transmission gain at $z\geq3 \lambda_0$. It should be noted that, the solution for the field coefficients presented in Eqs.~\eqref{eqa:coup_sol} and~\eqref{eqa} are very useful and provide a deep insight into the wave propagation inside the STM metasurface, especially for the luminal ST modulation (sonic regime), where the Bloch-Floquet-based analytical solution does not exist since the solution does not converge~\cite{Oliner_PIEEE_1963,Taravati_PRB_2017,Taravati_PRAp_2018}.
\indent Figure~\ref{Fig:24}(d) plots the result for the luminal STM metasurface in Fig.~\ref{Fig:24}(c), except for wave incidence from the right side (backward incidence). This figure shows that, in contrast to the forward wave incidence, here the incident wave passes through the metasurface with negligible alteration and minor transition of energy and momentum to the $m=-1$ STH.

\section{Nonreciprocal-Beam Metasurface}
Here, we present the concept, theoretical model and experimental implications of full-duplex nonreciprocal-beam-steering transmissive phase-gradient metasurfaces. Such metasurfaces are realized by exploiting the unique properties of the asymmetric frequency-phase electromagnetic transitions in time-modulated particles. It is shown that when the refractive-index of structures is  time-modulated, the incident wave experiences interband electromagnetic transitions between electromagnetic states due to temporal and spatial frequency shifts~\cite{winn1999interband,dong2008inducing,Taravati_PRB_2017,Taravati_Kishk_MicMag_2019}, in analogy to electronic transitions in semiconductors. Such metasurfaces may be placed on top of a source antenna, transforming their radiation pattern and providing different radiation patterns for the transmit and receive states. Such metasurfaces are composed of an array of twin time-modulated unit cells, each of which functioning four major operations, i.e. wave reception, nonreciprocal phase shift for nonreciprocal beam steering, filtering out of unwanted temporal harmonics, and wave radiation. 

\begin{figure}
	\begin{center}
		\includegraphics[width=1\columnwidth]{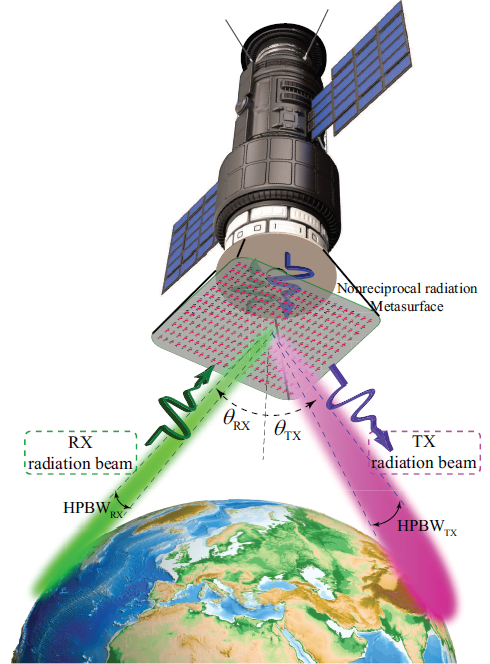}
		\caption{Full-duplex nonreciprocal radiation beam yielding highly directive reception and transmission radiation beams for efficient full-duplex point to point telecommunications~\cite{taravati2020full}.}
		\label{Fig:sat}
	\end{center}
\end{figure}

Figure~\ref{Fig:sat} illustrates the functionality of the nonreciprocal radiation beam from a gradient metasurface for efficient full-duplex point to point telecommunications. In the transmission state, the wave is launched from the source antenna traveling along the $+z$ direction, passes through the metasurface from region~1 to region~2 and radiates at an angle $\theta_\text{2,TX}$. In contrast, in the receive state, the metasurface presents the maximum transmission from region~2 to region~1 for the incoming wave at angle $\theta_\text{2,RX}$. Therefore, for a given radiation angle $\theta_0$, the metasurface is nonreciprocal, and may be represented by asymmetric and nonreciprocal radiation beams.

To realize the full-duplex nonreciprocal-beam-steering radome in Fig.~\ref{Fig:sat}, we consider a transmissive metasurface formed by an array of unit cells. In the transmission state, a plane wave with frequency $\omega_\text{i}$ impinges on the metasurface from the bottom left side with an angle of incidence $\theta_\text{1,TX}$. The
outgoing wave at $\theta_\text{2,TX}$ acquires a discrete phase-profile $\phi (md)=\phi_m$ where $m$ is the modulation phase at the $m$th unit cell and $d$ is the spacing between each two adjacent unit cells. However, in the transmission state, a plane wave with frequency $\omega_\text{i}$ impinges on the metasurface from the top right side with an angle of incidence $\theta_\text{2,RX}$. In contrast to the transmission state, and due to the conservation of momentum, the
outgoing wave at $\theta_\text{1,RX}$ acquires a discrete phase-profile $\phi(x)$, where $-\phi (md)=-\phi_m$. Assuming a constant gradient phase shift along the metasurface, the generalized Snell’s law of refraction yields
\begin{equation}\label{eq:tx}
\frac{\partial \phi_\text{TX}}{\partial x}=k_2 \sin(\theta_\text{2,TX}) -k_1\sin(\theta_\text{1,TX}),
\end{equation}
for the transmission state, and
\begin{equation}\label{eq:rx}
\frac{\partial \phi_\text{RX}}{\partial x}=k_2\sin(\theta_\text{2,RX})-k_1 \sin(\theta_\text{1,RX}) ,
\end{equation}
for the reception state. Here, $k_\text{1}$ and $k_\text{2}$ are the wave numbers in region 1 and region 2, respectively. Considering a constant phase gradient $\partial \phi_\text{MS}/\partial x$, the outgoing wave acquires anomalous refraction with respect to the incident
wave, whereas a spatially variant gradient, i.e, $\partial \phi_\text{MS}/\partial x$, leads to arbitrary radiation beams which enables beam-forming and advanced beam steering purposes.

\begin{figure}
	\begin{center}
\includegraphics[width=1\columnwidth]{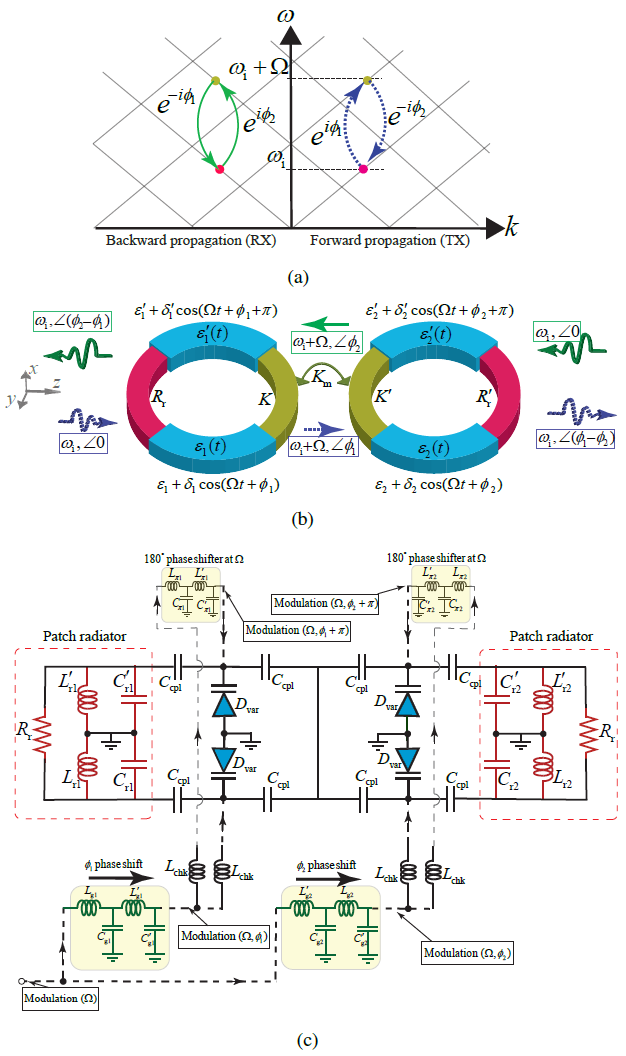}
		\caption{Nonreciprocal radiation based on electromagnetic transitions in time-modulated meta-atoms. (a) Qualitative dispersion diagram~\cite{taravati2020full}. (b) Architecture of the coupled time-modulated radiating loops introducing nonreciprocal phase shift~\cite{taravati2020full}. (c) Equivalent circuit model for the coupled meta-atoms architecture in (b)~\cite{taravati2020full}.}
		\label{Fig:26}
	\end{center}
\end{figure}

Figure~\ref{Fig:26}(a) shows a qualitative dispersion diagram of the structure for forward (green arrows) and backward (blue arrows) wave incidences, showing the general concept of the nonreciprocal phase shifting based on electromagnetic transitions in a time-modulated unit cell supporting two resonant frequencies, $\omega_\text{i}$ and $\omega_\text{i}+\Omega$. In the up-conversion, i.e., the electromagnetic transition from $\omega_\text{i}$ to $\omega_\text{i}+\Omega$, a phase shift of $\phi$ is achieved, whereas in the down-conversion, that is, the electromagnetic transition from $\omega_\text{i}+\Omega$ to $\omega_\text{i}$, a phase shift of $-\phi$ is introduced by the time modulation. 

The structure of the twin time-modulated unit cells is formed by four resonators, with electric permitivitties. In Fig.~\ref{Fig:26}(b) and~\ref{Fig:26}(c), $R_\text{r}$ and $R'_\text{r}$ represent the radiation resistances of the first and second unit cells, and $K$ and $K'$ denote the coupling between the arms of the first and second unit cells, whereas $K_\text{m}$ is the coupling between the two unit cells. In the forward incidence (left to right), the first time-modulated unit cell, characterized with permitivitties $\epsilon_1(t)$ and $\epsilon'_1(t)$, provides a frequency-phase transition from ($\omega_\text{i}$, $0$) to ($\omega_\text{i}+\Omega$, $\phi_1$). Then, the second time-modulated unit cell, characterized with permitivitties $\epsilon_2(t)$ and $\epsilon'_2(t)$, introduces a frequency-phase transition from ($\omega_\text{i}+\Omega$, $\phi_1$) to ($\omega_\text{i}$, $\phi_1-\phi_2$). In contrast, in the backward incidence (right to left), the second time-modulated unit cell, characterized with permitivitties $\epsilon_2(t)$ and $\epsilon'_2(t)$, provides frequency-phase transition from ($\omega_\text{i}$, $0$) to ($\omega_\text{i}+\Omega$, $\phi_2$), and then, the first time-modulated unit cell, characterized with permitivitties $\epsilon_1(t)$ and $\epsilon'_1(t)$, provides frequency-phase transition ($\omega_\text{i}+\Omega$, $\phi_2$) to ($\omega_\text{i}$, $\phi_2-\phi_1$). As a result, no frequency alteration occurs for both forward and backward transmitted waves, whereas a nonreciprocal phase shift is achieved, i.e., the backward transmitted wave acquires the phase shift of $\phi_2-\phi_1$ which is opposite to that of the forward transmitted wave phase $\phi_1-\phi_2$.

\begin{figure*}
	\begin{center}
\includegraphics[width=2\columnwidth]{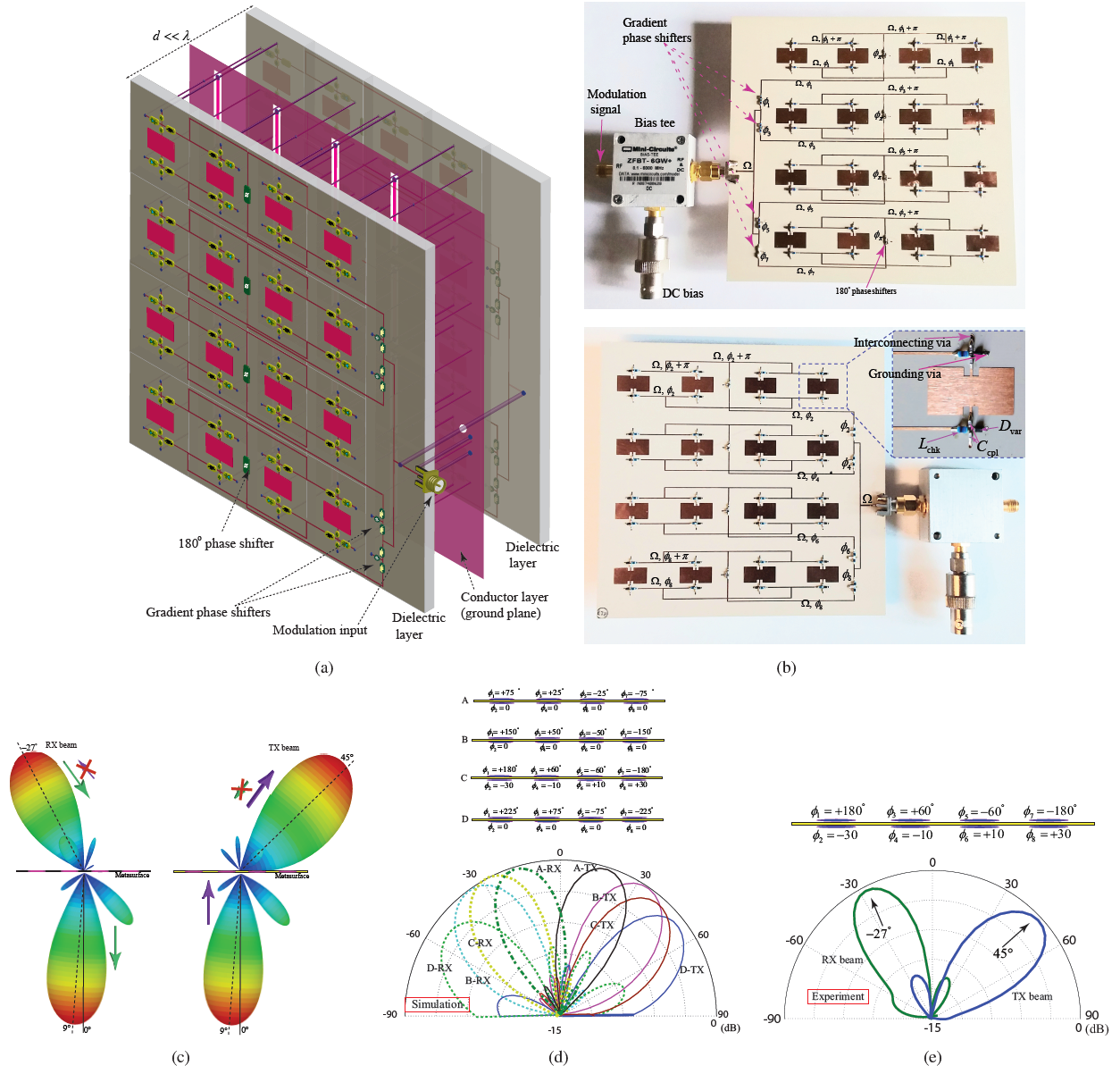}
		\caption{Nonreciprocal-beam-steering transmissive phase-gradient metasurface. (a) Schematic view~\cite{taravati2020full}. (b) The top and bottom views of the fabricated prototype, respectively~\cite{taravati2020full}. (c) and (d)	Full-wave simulation results~\cite{taravati2020full}. (e) Experimental results~\cite{taravati2020full}.}
		\label{Fig:27}
	\end{center}
\end{figure*}

Figure~\ref{Fig:27}(a) and~\ref{Fig:27}(b) show, respectively, a schematic and photo of the fabricated phase-gradient metasurface comprising $4 \times 4$ twin time-modulated unit cells.. The modulation signal is fed to the metasurface via a SMA connector. The metasurface includes eight gradient phase shifters ($\phi_1$ to $\phi_8$), i.e. four phase shifters in each side, providing the required modulation phase shifts for nonreciprocal beam steering. In addition, eight $180^\circ$ phase shifters ($\phi_\pi$) are utilized for achieving the $\pi$ phase-shifted version of each gradient phase shifted modulation signal. 

Figures~\ref{Fig:27}(c) to~\ref{Fig:27}(e) plot the full-wave simulation and experimental results for the nonreciprocal angle-symmetric/asymmetric transmission and reception radiation patterns of the nonreciprocal radiation beam metasurface for different angles of transmission and reception. The experimental isolation
between the transmission and reception radiation patters at specified transmission radiation angle ($\theta_\text{2,TX}=45^\circ$) is about $15.8$ dB, and the isolation at specified reception radiation angle ($\theta_\text{2,RX}=-27^\circ$) is about $10.4$ dB. To achieve higher isolation levels, one may change the modulation parameters or use a more directive metasurface by increasing the number of twin unit cells. 

\vspace{5mm}
\section{Antenna-Mixer-Amplifier Transceiver}

\begin{figure*}
\includegraphics[width=2\columnwidth]{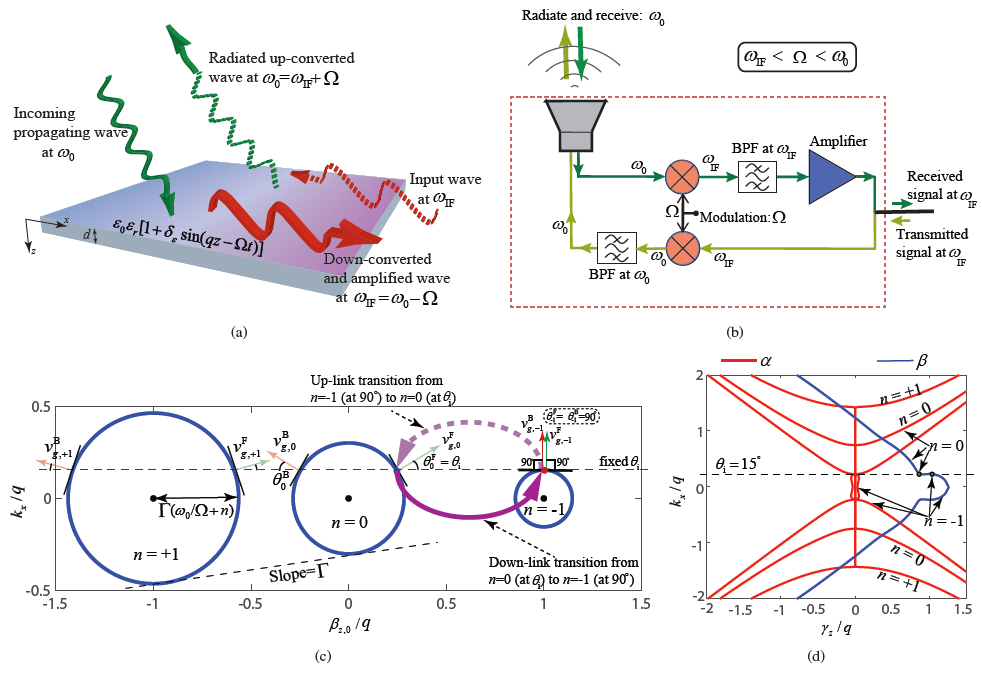}
	\caption{Antenna-mixer-amplifier metasurface. (a) Schematic representation showing the down-link and up-link wave transformations~\cite{Taravati_AMA_PRApp_2020}. (b) Circuital representation~\cite{Taravati_AMA_PRApp_2020}. (c) Analytical isofrequency dispersion diagram (for $\varGamma=0.2$ and $\epsilon_\text{m}=\rightarrow 0$) depicting down-link (reception) and up-link (transmission) electromagnetic transitions~\cite{Taravati_AMA_PRApp_2020}. (d) Analytical isofrequency dispersion diagram for $\varGamma=0.55$, including the real and imaginary parts of the $\gamma_{z,n}$, i.e., $\beta_{z,n}$ and $\alpha_{z,n}$, for $n=0$ and $n=-1$~\cite{Taravati_AMA_PRApp_2020}.
	} 
	\label{Fig:28}
\end{figure*}

We next show that a ST medium can function as a full transceiver front-end, that is, an antenna-mixer-amplifier-filter system. Specific related contributions of this study are as follows. Such an interesting functionality of the ST medium is endowed by ST surface waves. Other recently proposed ST antenna systems are formed by integration of the STM medium with an antenna~\cite{Taravati_APS_2015,Taravati_LWA_2017}, and hence suffer from a number of drawbacks, i.e., requiring long structures, low efficiency and narrow-band operation. The proposed antenna-mixer-amplifier introduces large frequency up- and -down conversion ratios. This is very practical, because in real-scenario wireless telecommunication systems, a large frequency conversion is required, i.e., a frequency conversion from a microwave/millimeter-wave frequency to an intermediate frequency in receivers. In contrast, recently proposed ST frequency converters suffer from very low frequency conversion ratios (up-/down-converted frequency is very closed to the input frequency)~\cite{Taravati_APS_2015,Taravati_LWA_2017,Grbic2019serrodyne}.

Consider the antenna-mixer-amplifier metasurface in Fig.~\ref{Fig:28}(a), with the thickness of $d$ and a STP electric permittivity. The metasurface in Fig.~\ref{Fig:28}(a)is obliquely illuminated by the $y$-polarized incident electric field given in Eq.~\eqref{eqa:Ei1}. As shown in Fig.~\ref{Fig:28}(a), in the receiving state (down-link), strong transition from a space-wave with temporal frequency $\omega_0$, to a ST surface wave with temporal frequency $\omega_\text{IF}=\omega_0-\Omega$, occurs.  In the transmission state (up-link), a transition from a ST surface wave at $\omega_\text{IF}$ to a space-wave at $\omega_\text{0}=\omega_\text{IF}+\Omega$ occurs.

Figure~\ref{Fig:28}(b) shows the functionality of the antenna-mixer-amplifier in Figs.~\ref{Fig:28}(a). Thanks to the unique properties of STM media, that will be described later in this study, such a medium introduces the functionality of a highly directive antenna, a pure frequency up-converter, a pure frequency down-converter, up-link and down-link filters, and a down-link amplifier. Such a rich functionality has not been experienced in other media unless several components and media are integrated together, as shown in Fig.~\ref{Fig:28}(b). However, here we show that a single medium can offer such versatile and useful operation.

As the metasurface is periodic in both space and time, the spatial and temporal frequencies of the ST harmonics inside the structure are governed by the momentum conservation law, i.e., $\gamma_{z,n}=k_{z,\text{i}}+ n q+i\alpha_{z,n}$ and the energy conservation law, i.e.,	$\omega_n=\omega_0+ n \Omega$. The incident angle reads $\theta_\text{i}=\sin^{-1} \left(1-\Omega/\omega_0 \right)$. In addition, to achieving a strong transition to the $n=-1$ harmonic, the scattered $n=-1$ harmonic inside the medium should propagate in parallel to the two ST surface waves on the two boundaries of the medium at $z=0$ and $z=d$, i.e., $\theta_{n=-1}=90^\circ$. Thus, using Eq.~\eqref{eqa:mod_angl}, we achieve $\beta_{z,-1}=0$. As a result, the $z$ component of the Wave vector inside the medium is purely imaginary, i.e., $\gamma_{z,-1}=i\alpha_{z,-1}$, whereas the incident field wavenumber $k_z$ is purely real.

The STM medium presents a transition from the fundamental harmonic $n=0$ to a large (theoretically infinite, $-\infty<n<\infty$) number of ST harmonics. Such a transition is very strong for the luminal ST modulation, where the ST modulation velocity is close to the background phase velocity, i.e., $v_\text{m} = v_\text{b}$~\cite{Taravati_PRB_2017,Taravati_PRAp_2018}. To prevent generation of strong undesired time harmonics, here the STM medium operates in the subluminal regime, where $0< v_\text{m}< v_\text{b}$, i.e., $0<\varGamma_{\text{subluminal}}<\sqrt{\epsilon_\text{r}/(\epsilon_\text{av}+\epsilon_\text{m})}$. As a result, a pure and precise transition between the fundamental ($n=0$) harmonic and the desired (here $n=-1$) ST surface wave harmonic can occur.

Figure~\ref{fig:10}(b) plots the analytical isofrequency dispersion diagram ($\beta_{z,n}(k_{x,\text{i}})$ at $\omega/\Omega=1.363$) of the sinusoidally ST surface wave medium with the electric permittivity in~\eqref{eqa:perm} for the subluminal regime, i.e., $\varGamma=0.55$ for $\epsilon_\text{m}\rightarrow 0$. It may be seen from this figure that forward $n=0$ and $n=-1$ harmonics are excited very close to each other, where $n=0$ is excited at an angle of scattering of $\theta_{0}=\theta_\text{i}=15^\circ$. However, the $n=-1$ harmonic is intentionally excited at the angle of scattering of $\theta_{-1}=90^\circ$. Figure~\ref{fig:10}(c) plots the same isofrequency $\beta_{z,n}(k_{x,\text{i}})$ diagram as Figure~\ref{fig:10}(b) except for a greater modulation amplitude of $\epsilon_\text{m}=0.45$. As a result of non-equilibrium in the electric and magnetic permitivitties of the medium, several electromagnetic badgaps appear at the intersections between ST harmonics~\cite{Taravati_PRAp_2018}. As a consequence, strong coupling between some of the harmonics has occurred, e.g. between the $n=0$ and $n=-1$ harmonics. 

Figure~\ref{Fig:28}(d) plots the complex isofrequency dispersion diagram $\gamma_{z,n}(k_{x,\text{i}})$ for the medium in Fig.~\ref{fig:10}(c). This diagram is formed by two different curves, i.e., the real $\beta_{z,n}(k_{x,\text{i}})$ and the imaginary $\alpha_{z,n}(k_{x,\text{i}})$ parts of the wavenumber. For the sake of clarification, we have included only a few number of harmonics. This figure shows that at the excited angle of incidence $\theta_\text{i}=15^\circ$, the $n=0$ harmonic introduces a pure real wavenumber, i.e., $\gamma_{z,0}=k_{z,\text{i}}$, whereas the $n=-1$ harmonic acquires a pure imaginary wavenumber, that is, $\gamma_{z,-1}=i\alpha_{z,-1}$. As a result, a perfect ST transition from a pure propagating wave to a pure ST surface wave is ensured.

\begin{figure*}
	\begin{center}
\includegraphics[width=2\columnwidth]{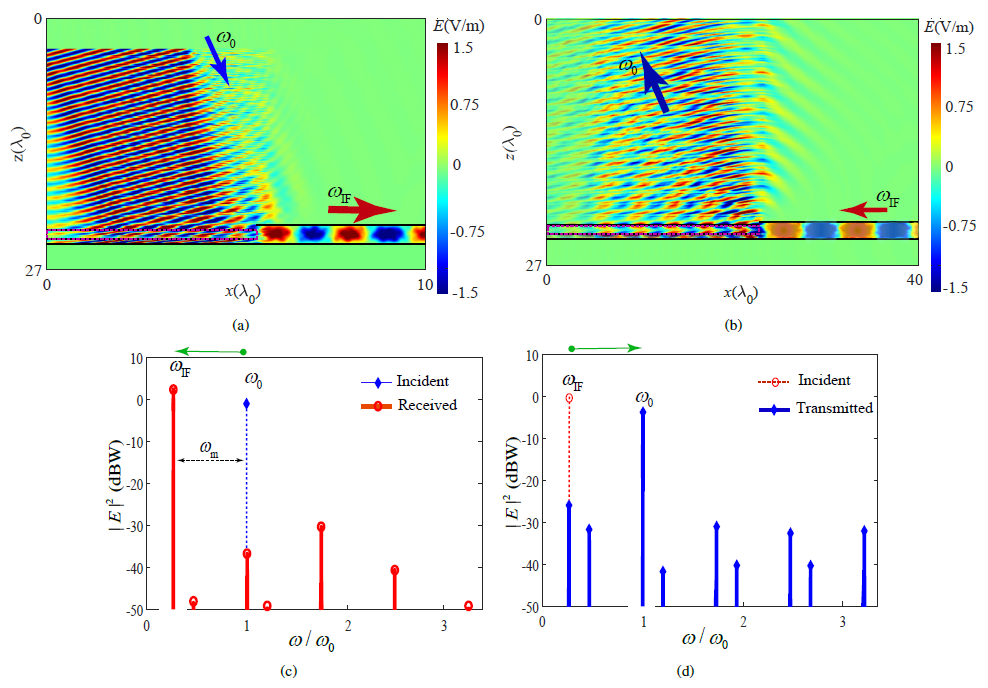}		
	\caption{FDTD simulation results of the antenna-mixer-amplifier ST surface wave medium with $\omega_0/\Omega=1.363$, $\epsilon_\text{m}=0.45$, $d=0.8 \lambda_0$, and $\theta_\text{i}==15^\circ$. (a) and (c) Time-domain and frequency-domain responses for the down-link transition~\cite{Taravati_AMA_PRApp_2020}. (b) and (d) Time-domain and frequency-domain responses for the up-link transition~\cite{Taravati_AMA_PRApp_2020}.}
	\label{Fig:29}
\end{center}
\end{figure*}

The metasurface assumes $\epsilon_\text{m}=0.45$, $\varGamma=0.55$, and $d=0.8 \lambda_0$. A plane wave with temporal frequency $\omega_0=1.363\Omega$ is propagating in the $+z$-direction under an angle of incidence of $\theta_\text{i}=15^\circ$, and impinges on the metasurface. Figure~\ref{Fig:29}(a) shows the time domain numerical simulation result for the receiving state (down-link). It may be seen from this figure that a pure transition from the incident space-wave at $\omega_0$ to the ST surface wave, propagating along the $x$-direction, at frequency $\omega_\text{IF}=\omega_\text{0}-\Omega$ occurs. Furthermore, it may be seen from Fig.~\ref{Fig:29}(a) that the amplitude of the received wave is stronger than the amplitude of the incident wave. Figure~\ref{Fig:29}(c) plots the frequency domain numerical simulation result. This plot clearly shows a pure and strong transition (frequency-conversion) from the incident wave to the down-converted ST surface wave. The 3.5 dB power conversion gain is in agreement with the analytical result. In addition, the amplitudes of the undesired harmonics are more than $33$ dB lower than the amplitude of the down-converted harmonic at $\omega_\text{IF}$.

Figure~\ref{Fig:29}(b) shows the time domain FDTD numerical simulation result for the transmission state (up-link). Here, a transition (up-conversion) from the ST surface wave at $\omega_\text{IF}$ to the space-wave at $\omega_\text{0}=\omega_\text{IF}+\Omega$ occurs. Figure~\ref{Fig:29}(d) plots the frequency domain numerical simulation result for the transmission state (up-link). This plot clearly shows a pure and strong transition (frequency-conversion) from the incident wave to the down-converted ST surface wave. The 3.52 dB power conversion loss is in agreement with the analytical result. In addition, the amplitude of the undesired harmonics are more than $27$ dB lower than the amplitude of the down-converted harmonic at $\omega_\text{0}$.

\newpage
\section{Conclusion}
We presented a comprehensive review of space-time metasurfaces. It is shown that space-time metasurfaces are capable of four-dimensional electromagnetic wave transformations which are significantly more versatile and useful than the three-dimensional wave transformations of conventional spatially variant static metamaterials and metasurfaces. We provided some of the unique applications of space-time metasurfaces, including nonreciprocal full-duplex wave transmission, pure frequency conversion, parametric wave amplification, spatiotemporal decomposition, space-time wave diffraction, and antenna-mixer-amplifier functionality. Recent progress on space-time metasurfaces for breaking time-reversal symmetry and reciprocity reveals a great potential for applications of such metasurfaces for low-energy and energy-harvesting telecommunication systems, and compact and integrated non-reciprocal devices, and sub-systems.

\newpage

\bibliographystyle{IEEEtran}
\bibliography{Taravati_Reference}

\end{document}